\documentclass[hyper,a4paper,notoc,11pt]{JHEP3}
\usepackage{amssymb,amsfonts,amsmath,graphicx,epsfig}
\usepackage{bm}% bold math f
\usepackage{epstopdf}
\newcommand\fverb{\setbox\fverbbox=\hbox\bgroup\verb}
\newcommand\fverbdo{\egroup\medskip\noindent%
			\fbox{\unhbox\fverbbox}\ }
\newcommand\fverbit{\egroup\item[\fbox{\unhbox\fverbbox}]}
\newbox\fverbbox

\title{Neutrino Mass, Sneutrino Dark Matter and Signals of Lepton Flavor Violation in the MRSSM}
\author{Abhishek Kumar$^{a}$, David Tucker-Smith$^{b}$ and Neal Weiner$^{a}$ \\ $ ^{a}$ Center for Cosmology and Particle Physics,
  Department of Physics, New York University,
New York, NY 10003, USA
\\ $^{b}$ Department of Physics, Williams College,
Williamstown, MA 01267, USA}
\abstract{We study the phenomenology of mixed-sneutrino dark matter in the Minimal $R$-symmetric Supersymmetric Standard Model (MRSSM).  Mixed sneutrinos fit naturally within the MRSSM, as the smallness (or absence) of neutrino Yukawa couplings singles out sneutrino $A$-terms as the only ones not automatically forbidden by $R$-symmetry.  We perform a study of randomly generated  sneutrino mass matrices and find that
({\em i})  the measured value of $\Omega_{DM}$ is well within the range of  typical values obtained for the relic abundance of the lightest sneutrino,
({\em ii}) with small lepton-number-violating mass terms $m_{nn}^{2} {\tilde n} {\tilde n}$ for the right-handed sneutrinos, random matrices satisfying the $\Omega_{DM}$ constraint have a decent probability of satisfying direct detection constraints, and much of the remaining parameter space will be probed by upcoming experiments,
({\em iii}) the   $m_{nn}^{2} {\tilde n} {\tilde n}$ terms  radiatively generate appropriately small Majorana neutrino masses, with neutrino oscillation data favoring a mostly sterile lightest sneutrino with a dominantly $\mu$/$\tau$-flavored active component, and
({\em iv})   a sneutrino LSP with a significant $\mu$ component can lead to striking signals of $e$-$\mu$ flavor violation in dilepton invariant-mass distributions at the LHC.
}
\begin{document}
\newcommand{\gsim}{ \mathop{}_{\textstyle \sim}^{\textstyle >} }
\newcommand{\be}{\begin{equation}}
\newcommand{\ee}{\end{equation}}
\newcommand{\ben}{\begin{equation*}}
\newcommand{\een}{\end{equation*}}
\newcommand{\bea}{\begin{eqnarray*}}
\newcommand{\eea}{\end{eqnarray*}}
\newcommand{\p}{\partial}
\newcommand{\tev}{{\rm TeV}}
\newcommand{\gev}{{\rm GeV}}
\newcommand{\kev}{{\rm keV}}
\newcommand{\neut}[2]{\tilde \chi^{0}_{#1,#2}}
\newcommand{\charg}[1]{\tilde \chi^{+}_{#1}}
\newcommand{\n}{\newline}
\newcommand{\snu}[1]{\tilde \nu_{#1}}
\newcommand{\vev}[1]{ \left\langle {#1} \right\rangle }
\newcommand{\til}{\tilde}
\newcommand{\xra}{\xrightarrow}
\newcommand{\ra}{\rightarrow}
\baselineskip=16pt
\def\Z{{\bf Z}}

\section{Introduction}
The weak scale promises to hold a wealth of new physics. Both the hierarchy problem and the success of WIMP dark matter candidates suggest that new particles may be discovered at the Tevatron or LHC. Indeed,
one of the appealing features of supersymmetric theories with R-parity is that the lightest superpartner (LSP) is a natural dark matter candidate.

At the same time, the minimal supersymmetric standard model (MSSM) is fraught with issues that invite us to explore new and interesting structures. In particular, the flavor problem of the MSSM is significant \cite{Gabbiani:1988rb,Gabbiani:1996hi}, requiring a suppression of flavor violating terms at the $10^{-3}$ level, even with CP conservation.

The Minimal $R$-symmetric Supersymmetric Standard Model (MRSSM) $\cite{Kribs:2007ac}$ was proposed as a new solution to the supersymmetric flavor problem without flavor-blind mediation. By imposing a continuous or suitably large discrete $R$-symmetry on weak scale supersymmetry, order one flavor violating soft masses for the sfermions are allowed.
The $R$-symmetry forbids Majorana gaugino masses, trilinear $A$-terms and the $\mu$-term. Since massless gauginos and Higgsinos are in conflict with experiment, the MSSM is augmented by considering $R$-symmetry preserving Dirac gauginos whose masses are generated by ``supersoft" operators $\cite{Fox:2002bu}$. Although previous attempts to implement Dirac gauginos $\cite{Hall:1990hq,Randall:1992cq,Dine:1992yw}$ were within the context of flavor-blind SUSY breaking, in the MRSSM they are considered  part of a general softly broken supersymmetric theory.

The phenomenology of the MRSSM is striking. Because the flavor violation is order one, dramatic flavor-violating signatures  $\cite{ArkaniHamed:1996au,Bityukov:1997ck,Agashe:1999bm,Bartl:2005yy}$ are possible. The TeV-scale Dirac gauginos and $\mathcal O(1)$ flavor violation in the sfermion sector open up the possibility of a natural mixed sneutrino dark matter candidate $\cite{ArkaniHamed:2000bq,Thomas:2007bu}$ as well as interesting flavor violating signals at the LHC $\cite{Kribs:2009zy}$.

The MSSM sneutrino was long ago considered an intriguing dark matter candidate $\cite{Hagelin:1984wv,Ibanez:1983kw}$ but
is today ruled out by the combined relic abundance, direct-detection, and invisible $Z$-width constraints.
By mixing the sneutrino with a gauge-singlet field,
the coupling of the lightest mass eigenstate to the $Z$ can be suppressed  and the invisible $Z$-width constraint is met for mixing angles satisifying $\sin\theta \leq$ 0.4 $\cite{ArkaniHamed:2000bq,Borzumati:2000mc,Chou:2000cy}$, for arbitrarily small masses. The appropriate relic abundance can also be achieved since the sneutrino annihilation rate in the early universe is suppressed $\cite{ArkaniHamed:2000bq}$. Related scenarios of sneutrino dark matter are explored in Refs.~ $\cite{Asaka:2005cn,Asaka:2007zz,Gopalakrishna:2006kr,Lee:2007mt,Arina:2007tm}$.

However, mixed-sneutrino dark matter is still tightly constrained by direct detection experiments\cite{ArkaniHamed:2000bq}, even with the suppression of the $Z$-coupling. Splitting the scalar and pseudo-scalar components can relax  direct-detection constraints \cite{Han:1997wn,Hall:1997ah,ArkaniHamed:2000bq}, but once experimental upper bounds on the radiatively generated neutrino masses are taken into account, the viable parameter space is rather finely tuned \cite{Arina:2007tm,Thomas:2007bu}. This makes it challenging to realize a model of sneutrino inelastic dark matter (iDM) \cite{TuckerSmith:2001hy} in standard supersymmetric scenarios.

These tensions are resolved in the MRSSM, where mixed-sneutrino dark matter moreover finds  a natural home.  Because the flavor violation is naturally large, there is a natural understanding of the origin of the large neutrino mixing angles through an anarchic \cite{Hall:1999sn} setup. Because the gauginos are Dirac, with only small Majorana masses coming from the conformal anomaly, the radiative masses are naturally a loop factor lower than in simple extensions of the MSSM, and are thus appropriately small, even with large enough lepton-violation to evade direct detection constraints. Finally, since neutrinos have small (or zero) Yukawas, they are singled out as the only superfields in the MRSSM to have large A-terms, which are essential for radiatively generating the neutrino mass matrix.

A study of mixed sneutrino dark matter in the MRSSM is thus  strongly motivated.
In this paper, we will explore the degree to which the anarchic scenario, implemented here, can produce neutrino oscillation parameters consistent with experimental results, while simultaneously satisfying direct detection constraints and yielding the appropriate relic abundance. The connection to neutrino masses will give us insight into the likely  flavor content of the sneutrino LSP,  which  will in turn motivate  analyses sensitive to lepton flavor violation at the LHC. Because the radiative corrections to neutrino masses are suppressed due to the Dirac gauginos, mixed-sneutrino iDM becomes viable again. If the scenario we explore is correct, we may learn that the one form of non-baryonic dark matter that we already know  (neutrinos) is intimately linked to the other (cold dark matter).

\section{Mixed sneutrinos in the MRSSM}
In addition to the elimination of $A$-terms, the $\mu$-term and Majorana gaugino masses, the MRSSM has additional fields that distinguish it from the MSSM. Adjoint chiral superfields are added to marry the standard gauginos. The Higgs sector is extended by adding multiplets $R_{u}$ and $R_{d}$ with the appropriate charges to allow $R$-symmetric $\mu$-terms to be written down with $H_{u}$ and $H_{d}$ respectively. The scalar components of the Higgs (and not the $R$-fields) acquire vevs that break electroweak symmetry, thereby preserving the $R$-symmetry.

Because the gauginos are Dirac, they can naturally be much heavier than the scalars \cite{Fox:2002bu}. The supersymmetric flavor problems are addressed by the TeV-scale Dirac gauginos and absence of $A$-terms at small tan $\beta$ and by the extended Higgs sector at large ${\rm tan} \beta$. Flavor violating observables in $\Delta F = 2$ processes are suppressed due to two effects. Radiative corrections to squark masses from Dirac gauginos are a {\em finite} one-loop effect, which allows them to be naturally lighter than in the MSSM, and the heavy gauginos provide a suppression by a factor $m_{\til q}^{2}/m_{\til g}^{2} \sim 10^{-2}.$ The $R$-symmetry allows leading flavor violating operators $\cite{Gabbiani:1988rb,Gabbiani:1996hi,Altmannshofer:2009ne}$ to be dimension 6 instead of 5 in the case of Majorana gauginos, thereby leading to another suppression $m_{\til q}^{2}/m_{\til g}^{2}\sim 10^{-2}.$ $\Delta F = 1$ processes such as $\mu \rightarrow e\gamma$ or $b \rightarrow s \gamma$, involve a helicity flip in the diagram. For Dirac gauginos, the opposite helicity state has no direct couplings to the sfermions and these diagrams are absent $\cite{Nelson:2002ca}$. Only the much smaller diagram with the helicity flip on the external fermion line or internal line from Higgsino-gaugino mixing can be present. In the MSSM, the radiatively generated couplings of down-type quarks to $H_{u}$ at large tan $\beta$ give the largest contributions to FCNCs $\cite{Hamzaoui:1998nu}$.  However, these contributions are absent in the MRSSM due to different diagrams being eliminated by the absence of $A$-terms, the $\mu$-term and Majorana gauginos.

Mixed sneutrino models arise naturally within the MRSSM. Since the holomorphic structure of the Yukawa and $A$-terms is the same, $R$-symmetry allows {\em either} Yukawa terms or $A$-terms to be written down, but not both. The electrically charged fermions of the Standard Model have large Yukawa couplings, and thus no $A$-terms are possible. In the neutrino sector, the Yukawas are zero, and thus large $A$-terms are natural.
The  trilinear interaction  $A_{ij} \til n_{i} \til l_{j} h_{u}$ couples
the left-handed sneutrinos to one or more additional singlet superfields $N_{i}$.
The large flavor violation in the slepton sector leads to large flavor mixing among the left-handed sneutrinos. By including three singlet superfields $N_{i}$, we obtain a $6 \times 6$ sneutrino mass matrix  with maximal flavor mixing and large off-diagonal $A$-terms that mix the left and right handed sneutrinos.

The inclusion of the $A$ terms, restricting to one generation, leads to the mass matrix
\be m_{\til \nu}^{2} = \begin{pmatrix} m_{L}^{2} + \frac{1}{2}m_{Z}^{2} \cos 2\beta  & A v \sin \beta \cr
A v \sin \beta  & m_{R}^{2}
\end{pmatrix}
\ee
where $m_{L}$ is the soft scalar mass for the left-handed sleptons, $\frac{1}{2} m_{Z}^{2} \cos 2\beta $ is the D-term contribution to the left-handed sneutrino mass-squared and $m_{R}$ is the soft scalar mass for the gauge-singlet right-handed sneutrino.

Although $m_{L}$,$m_{R}$ and $A$ are independent parameters, it is natural to have $m_{R} < m_{L}$ in the low-energy theory $\cite{ArkaniHamed:2000bq}$. Since $\til n$ is a standard-model singlet, there are no gaugino loops to drive its mass upward as the energy scale runs down from $M_{Pl}$ to $M_{W}.$ There are new, sizeable loop diagrams arising from the $A$-terms $(A_{ij} \til l_{i}\til n_{j} h_{u} + A_{ij} \til \nu_{i}\til n_{j} h_{u} )$ which push the soft masses down. Since two states, $\til l$ and $\til \nu,$ propagate in the loop contributing to $m_{R}$ and only one, $\til n,$ propagates in the loop contributing to $m_{L},$ $m_{R}$ is driven down faster than $m_{L}.$

\subsection{Radiative neutrino mass}

The $\Delta L$=2 Majorana neutrino mass term, $H_{u}H_{u}L_{L}L_{L}$ is $R$-symmetric and allowed in the MRSSM $\cite{Kribs:2007ac}$. In the presence of a lepton-number-violating mass term $m_{nn}^{2} \til n \til n$ for the right-handed sneutrinos, radiative neutrino mass can be generated via neutralino/sneutrino loops $\cite{ArkaniHamed:2000kj,Borzumati:2000mc,Borzumati:2000ya}$\footnote{To generate a $m_{nn}^{2} \til n \til n$ term we need $R$-symmetry breaking to feed into the right-handed sneutrino masses.  This gives us reason to expect this mass to be small  -- in the present framework we want it to be of order 100 MeV.}. These diagrams require a Majorana mass insertion and are absent if the gauginos are Dirac. In the MRSSM, anomaly mediation naturally generates a small Majorana gaugino mass in addition to the large, $\mathcal O$(TeV) Dirac mass. Furthermore, since the Dirac gauginos are somewhat heavier than in the MSSM, the radiative neutrino mass is suppressed and allows small neutrino masses without tuning.

The radiatively generated neutrino mass can be calculated in the mass insertion approximation $\cite{Hall:1985dx}$. Choosing a basis for the fermions and bosons where the gauge couplings to the neutralinos are flavor diagonal, the flavor changing is exhibited by non-diagonality of the sfermion propagators.
In the mass-eigenstate basis,
the neutrino mass diagram has odd numbers of insertions from the Majorana neutralino mass and from $\Delta_{ij},$ the lepton number violating mass term in the mass basis.
The Majorana neutralino mass can be treated as a perturbation on the Dirac neutralino mass matrix.

\begin{figure}[h]
\begin{center}
\includegraphics[width = 0.5\textwidth,viewport=76pt 470pt 593pt 785pt]{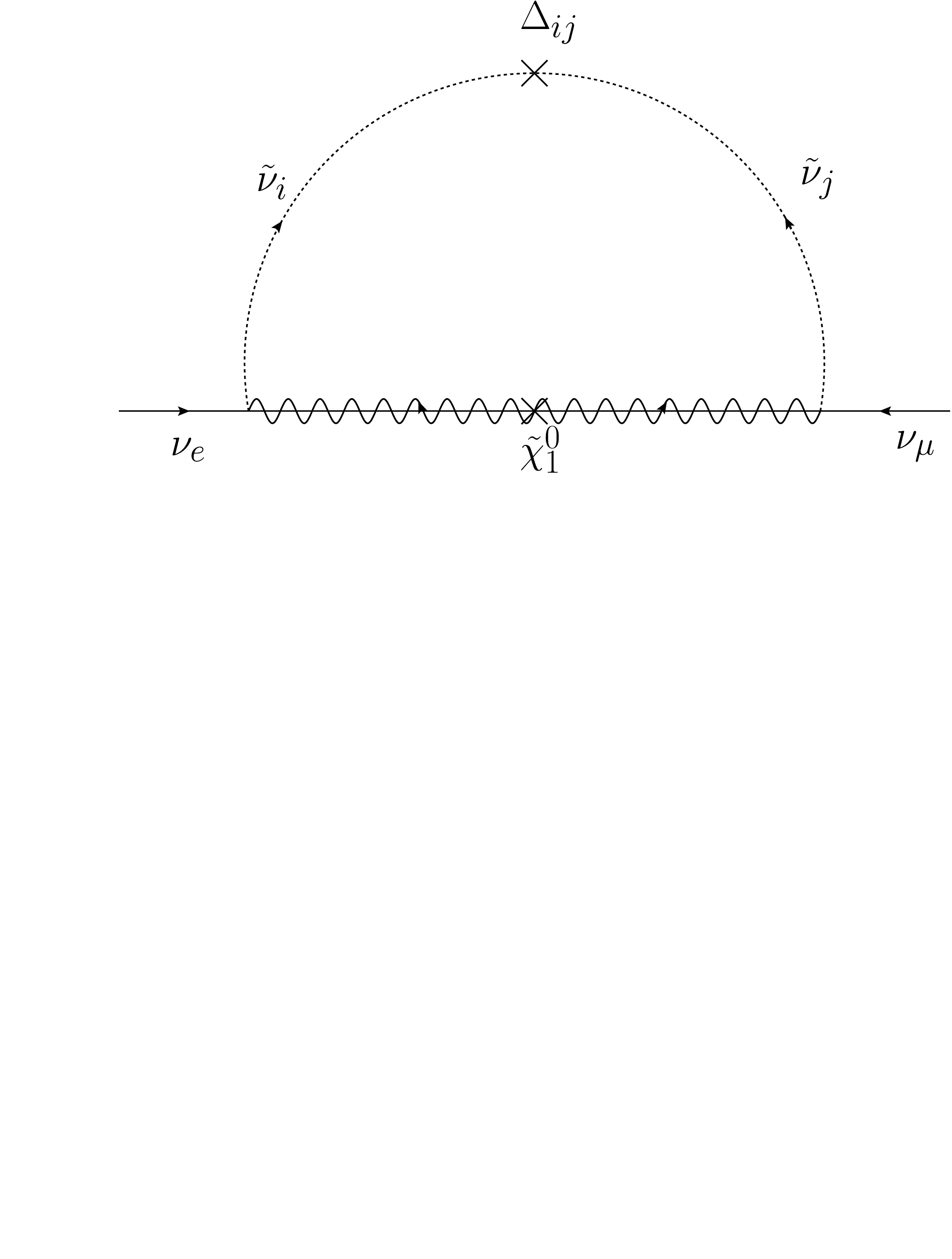}
\caption{The radiatively generated Majorana neutrino mass in the mass eigenstate basis}
\label{numass}
\end{center}
\end{figure}

For a single wino Majorana mass insertion $m_{\til W}^{M}$ ($m_{\til W}$ is the Dirac wino mass), the diagram is calculated as
(Fig. \ref{numass}):
\be m_{\nu_{e\mu}} = \sum_{i,j} I_{ij} \ee
where
\begin{align}
\label{eq:Iij1}
I_{ij}& = \int \frac{d^{4}p}{2\pi^{4}} g_{ei} \frac{1}{p^{2}-m_{i}^{2}} \Delta_{ij} \frac{1}{p^{2}-m_{j}^{2}} g_{j\mu}^{*} \frac{m_{\til W}^{M}}{p^{2}-m_{\til W}^{2}} \nonumber  \hskip 0.7in (i \ne j)\\
& = \frac{1}{32\pi^{2}} \frac{g_{ei} \Delta_{ij} g_{j\mu}^{*} m_{\til W}^{M}}{(m_{\til W}^{2}-m_{i}^{2})(m_{\til W}^{2}-m_{j}^{2})(m_{i}^{2}-m_{j}^{2})} \nonumber \\ &\qquad \left(m_{i}^{2} m_{\til W}^{2} \log[m_{\til W}^{2}/m_{i}^{2}] + m_{j}^{2} m_{\til W}^{2} \log[m_{j}^{2}/m_{\til W}^{2}]+ m_{i}^{2} m_{j}^{2} \log[m_{i}^{2}/m_{j}^{2}]\right), \\
\label{eq:Iij2}
I_{ii}& = \int \frac{d^{4}p}{2\pi^{4}} g_{ei} \frac{1}{p^{2}-m_{i}^{2}} \Delta_{ii} \frac{1}{p^{2}-m_{i}^{2}} g_{i\mu}^{*} \frac{m_{\til W}^{M}}{p^{2}-m_{\til W}^{2}} \nonumber \\ & = \frac{1}{32\pi^{2}} \frac{g_{ei} \Delta_{ii} g_{i\mu}^{*} m_{\til W}^{M}}{(m_{\til W}^{2}-m_{i}^{2})} \left(1+ \frac{m_{\til W}^{2}}{m_{\til W}^{2}-m_{i}^{2}} \log[m_{\til W}^{2}/m_{i}^{2}]\right),
\end{align}
and $i,j$ run over the 6 mixed sneutrinos.

The neutrino mass matrix thus obtained can be diagonalized with a matrix $U$.
Experimental results on neutrino masses and mixings can thus give  insight into the structure of the sneutrino mass matrix  $\cite{Hall:1999sn}$.

In our analysis below, we will consider experimental constraints on the following quantities from the neutrino sector:
\begin{align}
R & = \Delta m_{12}^{2}/\Delta m_{23}^{2} ,\\
s_{C} & = 4 |U_{e3}|^{2} (1-|U_{e3}|^{2}),\\
s_{atm} & = 4 |U_{\mu3}|^{2} (1-|U_{\mu3}|^{2}),\\
s_{\odot} & = 4|U_{e2}|^{2}|U_{e1}|^{2}, \end{align}
where $\Delta m_{12}^{2}$ is the smallest splitting and $\Delta m_{23}^{2}$ is the next largest splitting of the neutrino masses.

The current experimental bounds on these quantities (at 3 $\sigma$) are $\cite{Maltoni:2004ei}$:
\begin{align}
& 0.027\leq R \leq 0.04, \\
& s_{C} \leq 0.05,\\
& 0.34\leq s_{atm} \leq 0.67, \\
& 0.26\leq s_{\odot} \leq 0.4. \end{align}

\subsection{Direct detection}

Direct detection experiments constrain the spin-independent and spin-dependent WIMP-nucleon cross sections.
In the presence of lepton-number violation, the mixed LSP sneutrino ($\snu1$)  can evade direct detection constraints  arising from from $Z$ boson exchange $\cite{TuckerSmith:2001hy,TuckerSmith:2002af,Hall:1997ah}$.
However, the $h$-$\snu1$-$\snu1$ coupling is modified due to
the additional gauge degrees of freedom and also due to the large $A$-terms.
Therefore, scattering from Higgs exchange still constrains the theory $\cite{ArkaniHamed:2000bq}$. In the decoupling limit ($m_{A} \gg m_{h}$), the cross section for Higgs-mediated $\snu1$-nucleon scattering is
\be \sigma_{0} = \left(\frac{g_{hNN}}{1.26\times10^{-3}}\right)^{2}\left(\frac{g_{h\snu1\snu1}}{m_{N}+m_{\snu1}}\right)^{2}\left(\frac{115 \; \rm GeV}{m_{h}}\right)^{4}(2.48 \times 10^{-43} \rm cm^{2}),\ee
where $m_{N}$ is the nucleon mass,  we take $g_{hNN} = 1.26 \times 10^{-3}$ for the Higgs-nucleon coupling, with the up, down, strange and heavy quark contributions taken from $\cite{Cheng:1988im,Hatsuda:1990uw,Shifman:1978zn}$, and $g_{h\snu1\snu1}$ is the coupling of the light Higgs boson to the LSP sneutrino $\snu1.$ Recent lattice measurements of the light and strange quark condensates in the proton affect the cross-section evaluated here by $\sim$ 20$\%$ at most $\cite{Giedt:2009mr}$.

The $h$-$\snu1$-$\snu1$ coupling receives contributions from the Higgs $D$-term and the large $A$-terms. For pure Dirac gaugino masses, the $D$-term is zero and the Higgs quartic vanishes. Electroweak symmetry breaking can be recovered by adding  small Majorana masses that lift the $D$-flat direction. The resultant coupling for the LSP sneutrino in the mass basis becomes

\be g_{h\snu1\snu1} = \sum_{\kappa,\rho} \left(-\frac{1}{2} \left(\frac{m_{A_{2}}^{2}}{m_{A_{2}}^{2}+4 m_{D_{2}}^{2}} g_{2}^{2} + \frac{m_{A_{Y}}^{2}}{m_{A_{Y}}^{2}+4 m_{D_{Y}}^{2}} g_{Y}^{2} \right) v \cos 2\beta U_{1\kappa}^{2} - 2 \frac{1}{\sqrt{2}} A_{\kappa \rho} \sin \beta U_{1\kappa} U_{\rho 1} \right),
\label{equation:hsnucoupling}
\ee
  where $m_{A_{i}}$ and
 $m_{D_{i}}$ are the Majorana and Dirac gaugino masses,
 and $U$ is the mixing matrix that relates the sneutrino mass ($i$) and gauge ($\alpha$) eigenstates by $\tilde \nu_{i} = U_{i\alpha} \tilde \nu_{\alpha}.$ In the formula above, $\kappa$ runs over the active flavors ($e$,$\mu$,$\tau$) and $\rho$ runs over the three sterile flavors of  right-handed sneutrinos.

Currently, the strongest direct detection bounds on WIMP-nucleon scattering are from CDMS $\cite{Ahmed:2008eu}$ and XENON10 $\cite{Angle:2007uj}$. CDMS sets an upper limit on the spin-independent cross section of 4.6$\times$10$^{-44}$ cm$^{2}$ at the 90$\%$ confidence level for a WIMP mass of 60 GeV/$c^{2}$ and XENON10 has an upper bound of 4.5 $\times$ 10$^{-44}$ cm$^{2}$ for a WIMP mass of 30 GeV/$c^{2}.$

\subsection{Invisible $Z$-width}

The invisible $Z$-width is extremely constrained. From current experimental bounds, $\Gamma (Z\rightarrow \rm invisible)$ = 499.0 $\pm$ 1.5 MeV and $\Gamma (Z\rightarrow \nu\bar\nu)$ = 501.65 $\pm$ 0.11 MeV. Thus, new light degrees of freedom with a total mass less than $M_{Z}/2$ that couple to the $Z$ boson must have a decay rate within $\Delta \Gamma < 3$ MeV $\cite{Amsler:2008zzb}$.

The rate for $Z$ decaying to two mixed sneutrinos is given by
\be \Gamma_{\til \nu_{i}\til\nu_{j}} = \frac{1}{2} \Gamma_{\nu} \left(1-\frac{(m_{i}+m_{j})^{2}}{M_{Z}^{2}}\right)^{3/2} \left(1-\frac{(m_{i}-m_{j})^{2}}{M_{Z}^{2}}\right)^{3/2} U_{i\kappa}U_{j\kappa},\ee where $U$ is the mixing matrix that relates the mass and gauge sneutrino eigenstates, $\kappa$ runs over $e$,$\mu$,$\tau$ and $\Gamma_{\nu} $ = 167 MeV. The invisible $Z$-width constraint becomes \be \sum_{i,j} \Gamma_{\tilde \nu_{i} \tilde \nu_{j}} < \Delta \Gamma \ee where $m_{i}+m_{j} < M_{Z}.$

\subsection{Relic abundance}

The dominant annihilation channels for mixed sneutrinos in the early universe are $s$-channel $Z$-exchange, $t$-channel neutralino exchange and $s$-channel Higgs exchange. The Higgs-sneutrino coupling is enhanced due to large $A$-terms and the mixing suppresses the $Z$-exchange contribution so the $s$-channel Higgs exchange is often dominant. Sneutrino masses near the Higgs or $Z$ poles should give a depleted relic abundance due to rapid annihilation.

In the case of 6 mixed sneutrinos, the LSP sneutrino mainly contributes to the dark matter relic abundance. The heavier sneutrinos annihilate into the LSP sneutrino before freezeout. Co-annihilation is possible if the mass difference between two sneutrinos is small. The current bound on the dark matter energy density is 0.094 $\leq \Omega_{DM}h^{2} \leq $ 0.128 $\cite{Dunkley:2008ie}$.

\section{Analysis}

The sneutrino mass matrix in the MRSSM can have order one flavor mixing in the left and right handed sectors as well as large off-diagonal $A$-terms. The large flavor violation is shielded from and uncorrelated with the flavor observables.
The flavor violating terms and $A$-terms in the sneutrino mass matrix can be essentially $random$ without contributing to the supersymmetric flavor problem.

However, the sneutrino mass matrix is not unconstrained, due to limits from the radiatively generated neutrino masses and the invisible $Z$-width. As a plausible dark matter candidate, the sneutrino mass matrix is further constrained and must satisfy direct detection and relic abundance bounds, as discussed above.

We perform a systematic study of randomly generated $6\times6$ sneutrino mass matrices and analyze the likely properties of a mixed sneutrino dark matter candidate consistent with the experimental constraints discussed above. Our analysis sheds light on the mass hierarchies of the sneutrinos, whether the dark matter is likely to be predominantly active or predominantly sterile, and the expected flavor composition of the active part.

We generate 75 million random $6\times6$ Hermitian (real, symmetric) sneutrino mass matrices with positive eigenvalues. Each sneutrino matrix has $6$ random entries for $m_{L}$ and $m_{R}$, $9$ for the $A$-terms and $3$ for $\Delta,$ the lepton number violating mass terms for the right handed sneutrinos. The matrix entries are randomly sampled from a uniform distribution. The entries for $m_{L},m_{R}$ are sampled from $[0,250]$ GeV and the range for $A$ is $[-100,100]$ GeV. We take the elements of $\Delta$ to be  small perturbations on the sneutrino mass matrix, also with linear distributions. In our analysis the only effect of $\Delta$ is to generate neutrino masses, and only the ratios of the elements of $\Delta$ to one another are important.  This is because the neutrino masses are also directly proportional to the Majorana gaugino masses, which we also treat as a small perturbations.
Two cases are considered:  Run I, with heavy gauginos and an additional loop-generated flavor diagonal mass of $200$ GeV for the left handed sneutrinos (arising from gauge mediation), and Run II with light gauginos and without the flavor diagonal mass.

To calculate the relic abundance of the mixed sneutrino LSP, we use the MicrOmegas 2.0.7 code $\cite{Belanger:2007zz}$. The randomly generated sneutrino matrix is related to a random slepton matrix by the left-handed slepton and sneutrino mass difference arising from the $D$-term contribution. Thus the left-handed sleptons are also mixed in this analysis. The relevant couplings of the lightest 4 mixed sneutrinos and the 3 left-handed sleptons are modified by the appropriate mixing angles in the MSSM files in CalcHEP $\cite{Pukhov:2004ca}$ and fed into MicrOmegas to calculate the relic abundance. Since the lightest sneutrinos contribute to the relic abundance calculation, only the lightest 4 mixed sneutrinos are added to the code.

Although the right-handed slepton mass  matrix is also potentially  mixed, we take it to be proportional to the identity matrix in this analysis. The right-handed slepton couplings do not affect the relic abundance calculation significantly, and restricting flavor violation to the left-handed sleptons leads to a clearer analysis of the collider signals discussed in Section \ref{sec:collider}.

The gauginos are taken to have $\mathcal O$(TeV) masses.
As we have seen in equation \ref{equation:hsnucoupling}, the $h$-$\snu1$-$\snu1$ contains free parameters due to the modified Higgs $D$-term.
However, the contributions to this coupling from the $A$-terms are numerically more important, and so we can safely neglect the $D$-term contribution to the coupling.

We can now present the results of imposing  the various experimental constraints on the sneutrino mass matrices.

\section{Results}
\label{sec:results}

A table of the statistics obtained from applying neutrino mass, relic abundance, invisible $Z$-width and direct detection constraints is given below (Table \ref{table:Allstats}). We generate 75 million sets of matrices, first for the case with a flavor-diagonal mass of $200$ GeV (Run I - Heavy gauginos), and then for the case without  a  flavor-diagonal mass (Run II - Light gauginos).

 One interesting question is  whether what we know about neutrino masses and mixings makes the scenario more or less viable. To answer this question, we first calculate  the efficiency with which matrices passing the neutrino mass cuts subsequently satisfy the relic abundance, invisible $Z$-width and direct detection constraints.  We then compare this to the efficiency with which {\em random} sets of matrices -- ones not subjected to the neutrino mass cuts -- satisfy the same remaining constraints.    The  ``Random I'' and ``Random II'' rows of Table \ref{table:Allstats} show the results for 10,000 sets of  random matrices unconstrained by neutrino measurements, for comparison with Run I and Run II, respectively.

 The results in Table \ref{table:Allstats} show that matrices which satisfy the neutrino mass constraints are about as likely to satisfy the remaining constraints as  purely random matrices are.
Thus, fitting to neutrino mass properties does not seem to benefit or impede the search.

\begin{table}[h]
\centering
\begin{tabular}{|c|c|c|c|c|c|}
\hline
Run & $m_{\nu}$ & $m_{\nu}$+$\Omega_{DM} h^2$ & $m_{\nu}$+$\Omega_{DM} h^2$ & $m_{\nu}$+$\Omega_{DM} h^2$ & \% events \\
 & & & +$\Gamma_{Z\rightarrow \rm inv}$ & +$\Gamma_{Z\rightarrow \rm inv}$+direct & \\\hline
I & 7754 & 389 & 343 & 69 & 0.89  \\
Random I & 10000 & 428 & 369 & 106 & 1.06\\
II & 5757 & 273 & 144 & 12 & 0.21 \\
Random II & 10000 & 309 & 200 & 27 & 0.27\\
\hline
\end{tabular}
\caption{Results of experimental cuts on  random sneutrino mass matrices, as explained in the text.}
\label{table:Allstats}
\end{table}

\subsection{Neutrino Masses and Mixings}
About one in $10^{4}$ of the random sets of of sneutrino mass matrices satisfy the neutrino mass constraints.
Of course, this small efficiency is not by itself  a discouraging result, because as the measurement errors associated with the various experimental constraints become smaller and smaller, the fraction of random matrices passing all cuts will inevitably tend toward zero.
The question, then, is how {\em should} we evaluate the efficiency which which our matrices  pass the neutrino mass cuts?

Probably the simplest approach is to think of  the cuts as tunings. That is, $s_C < 0.05$ is a 5\% tuning, while the constraints on $s_\odot$, $s_{atm}$, and $R$ are  14\%,  33\%, and  1.3\% tunings. With this interpretation, we would expect only a fraction $\sim 3\times 10^{-5}$ of the total to pass the cuts. The $10^{-4}$ which pass is (marginally) better than this,  suggesting that there is  nothing particularly surprising about the degree to which neutrino measurements  constrain the sneutrino mass matrices.

That said, the fact that a particular region of parameter space {\em has} been singled out allows us to make more specific statements about the nature of the LSP, as we will see below.
\begin{figure}
\begin{center}
\includegraphics[width = 0.49\textwidth]{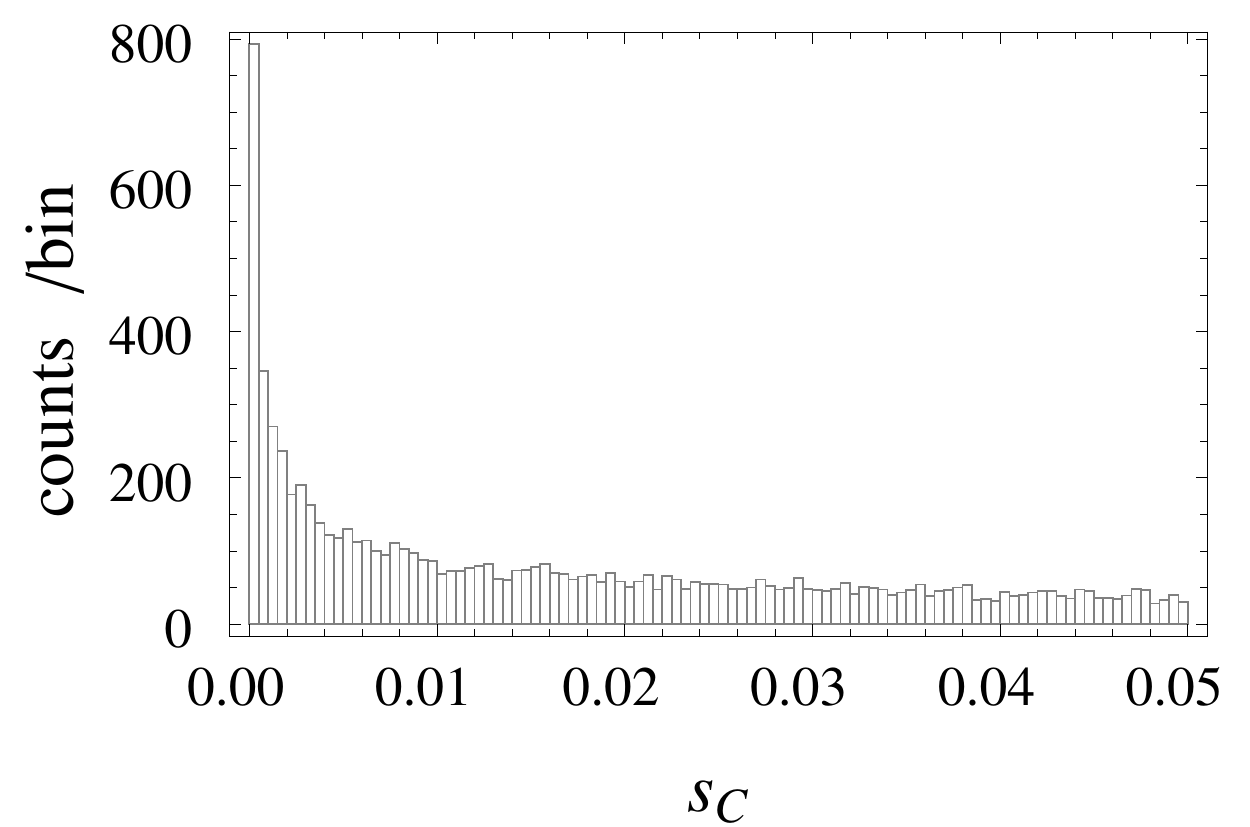}
\includegraphics[width = 0.49\textwidth]{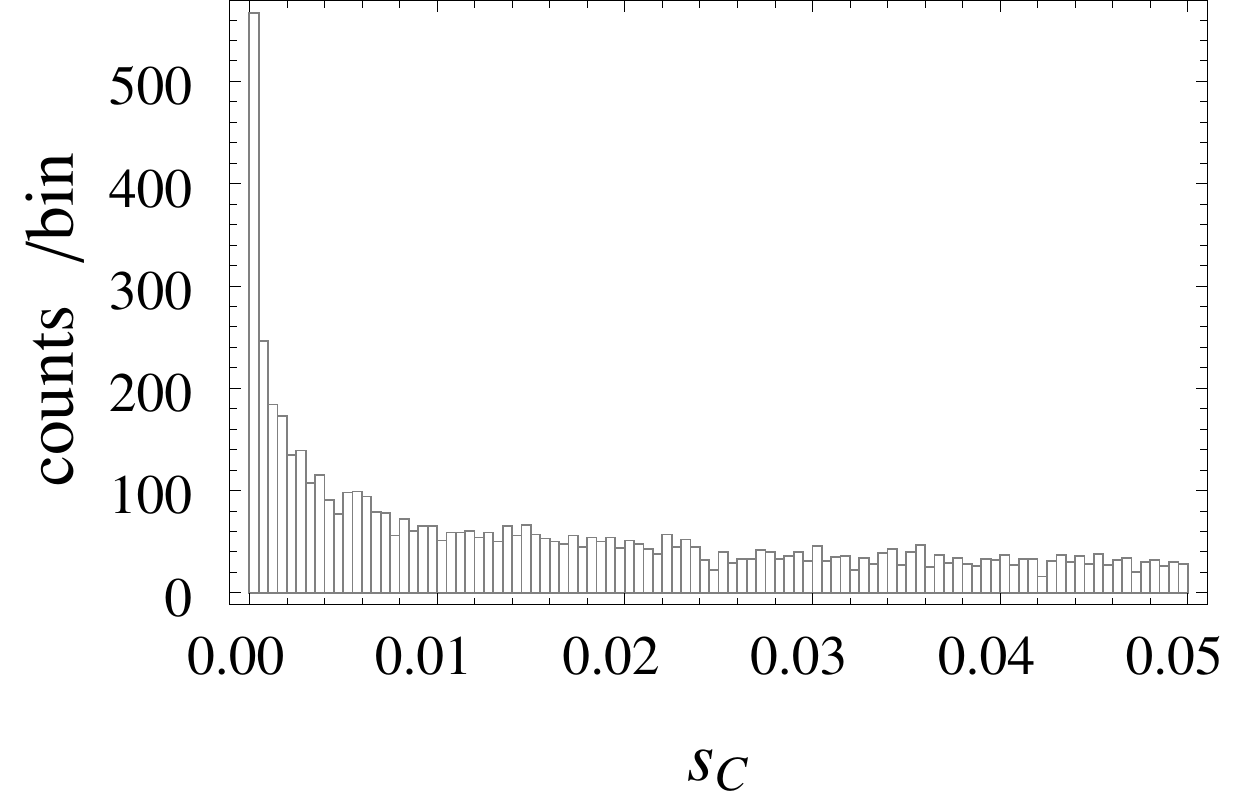}
\caption{The $s_{C}$ distributions peak towards zero for the samples that satisfy the neutrino cuts. (i) Run I (ii) Run II}
\label{theta13}
\end{center}
\end{figure}

Before we get to the LSP, let us consider the neutrino mass matrix in more detail.
The radiative Majorana neutrino mass matrix generated via neutralino/sneutrino loops is also random. A table of the statistics obtained for
the various neutrino mass and mixing cuts
is given above (Tables \ref{table:RunIstats} and \ref{table:RunIIstats}). The neutrino mass matrices that pass the neutrino cuts generally exhibit a normal mass hierarchy, with less than $0.5\%$ exhibiting an inverted hierarchy.

\begin{table}
\centering
\begin{tabular}{|l|c|c|c|c|}
\hline
Majorana & no cuts & $s_{atm}$ & $s_{\odot}$ & $s_{atm} + s_{\odot}$ \\ \hline
no cuts & 75,000,000 & 14,845,534  & 8,069,129 & 1,388,589  \\ \hline
$s_{C}$ & 9,007,887 & 1,633,398 & 760,902 & 142,528 \\ \hline
$R$ & 3,786,775 & 405,762 & 717,334 & 67,116 \\ \hline
$s_{C}+R$ & 490,465 & 84,996 & 43,169 & 7,754 \\ \hline
\end{tabular}
\caption{The impact of neutrino mass and mixing cuts for  Run I}
\label{table:RunIstats}
\end{table}

\begin{table}
\centering
\begin{tabular}{|l|c|c|c|c|}
\hline
Majorana & no cuts & $s_{atm}$ & $s_{\odot}$ & $s_{atm} + s_{\odot}$ \\ \hline
no cuts & 75,000,000 & 15,508,313  & 7,861,926 & 1,445,386 \\ \hline
$s_{C}$ & 9,488,149 & 1,750,349 & 816,193 & 159,378 \\ \hline
$R$ & 2,690,087 & 284,677 & 541,726 & 51,241 \\ \hline
$s_{C}+R$ & 353,406 & 62,412 & 31,165 & 5,757 \\ \hline
\end{tabular}
\caption{The impact of neutrino mass and mixing cuts for  Run II}
\label{table:RunIIstats}
\end{table}

To check that the neutrino masses obtained are in the experimentally viable range, we consider a set of benchmark values for a $2\times2$ sneutrino matrix (Table \ref{table:neutmass}). $\sin \theta$ is the mixing angle and $\delta$ is the mass splitting between the lightest CP-even and CP-odd sneutrino. A splitting of order $100\;\kev$ ensures that inelastic scattering at XENON10 is kinematically impossible. The Dirac Wino mass is heavy and the Majorana Wino mass is taken to be a one-loop factor down from the Dirac mass. In this case, the neutrino mass is given by $\cite{Thomas:2007bu}$

\be m_{\nu} = \frac{1}{32\pi^{2}} g^{2} \sin^{2}\theta \delta m_{\snu1} m_{\til W}^{M} \sum_{i,j} I_{ij}, \ee
where $I_{ij}$ is defined in \ref{eq:Iij1}, \ref{eq:Iij2}. The benchmark point gives $m_{\nu} = 0.05$ eV, consistent with experimental limits on the sum of neutrino masses $\sim 1$ eV. Furthermore, varying the Dirac Wino mass between $500\;\gev$ and $1\;\tev$ does not change the statistics from the neutrino mass and mixing cuts.

\begin{table}
\centering
\begin{minipage}{0.45\textwidth}
\centering
\begin{tabular}{|c|c|}
\hline
$ m_{\snu1}$  &  $100\;\gev$  \\
\hline
$ \rm sin\;\theta$ & $0.2$ \\
\hline
$ m_{\til W}^{D}$ & $1\;\tev$  \\
\hline
\end{tabular}
\end{minipage}
\begin{minipage}{0.45\textwidth}
\centering
\begin{tabular}{|c|c|}
\hline
$ m_{\snu2} $ & $150\;\gev$ \\
\hline
$ \delta $ & $100\;\kev$ \\
\hline
$ m_{\til W}^{M}$ & $m_{\til W}^{D}/16\pi^{2}$ \\
\hline
\end{tabular}
\end{minipage}
\caption{Benchmark point gives an experimentally viable neutrino mass, $m_{\nu}=0.05$ eV}
\label{table:neutmass}
\end{table}

Experimentally, the observable $s_{C}$ which corresponds to the $\theta_{13}$ angle in the neutrino mixing matrix $U$ is the least constrained $\cite{Maltoni:2004ei}$. The randomly generated sample of neutrino mass matrices tends to have a $s_{C}$ distribution that peaks towards zero mixing, as shown in Fig. \ref{theta13}. A non-zero measurement of $\theta_{13}$ at Daya Bay or at another similar experiment, for instance, would support an anarchic scenario, but 
even if $\theta_{13}$ were constrained to be smaller than 0.01, that would not exclude it.

\begin{figure}[h]
\begin{center}
\includegraphics[width = 0.49\textwidth]{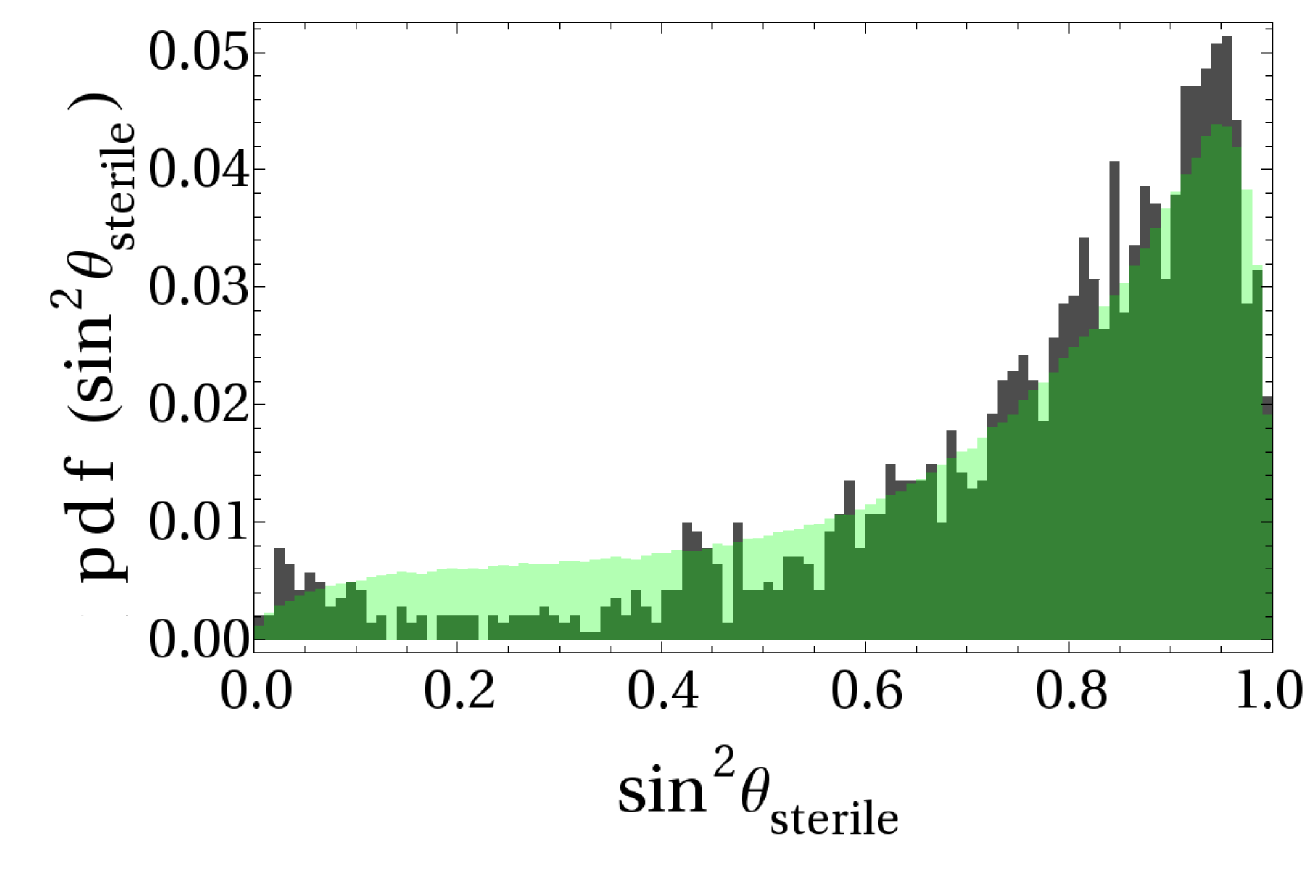}
\includegraphics[width = 0.49\textwidth]{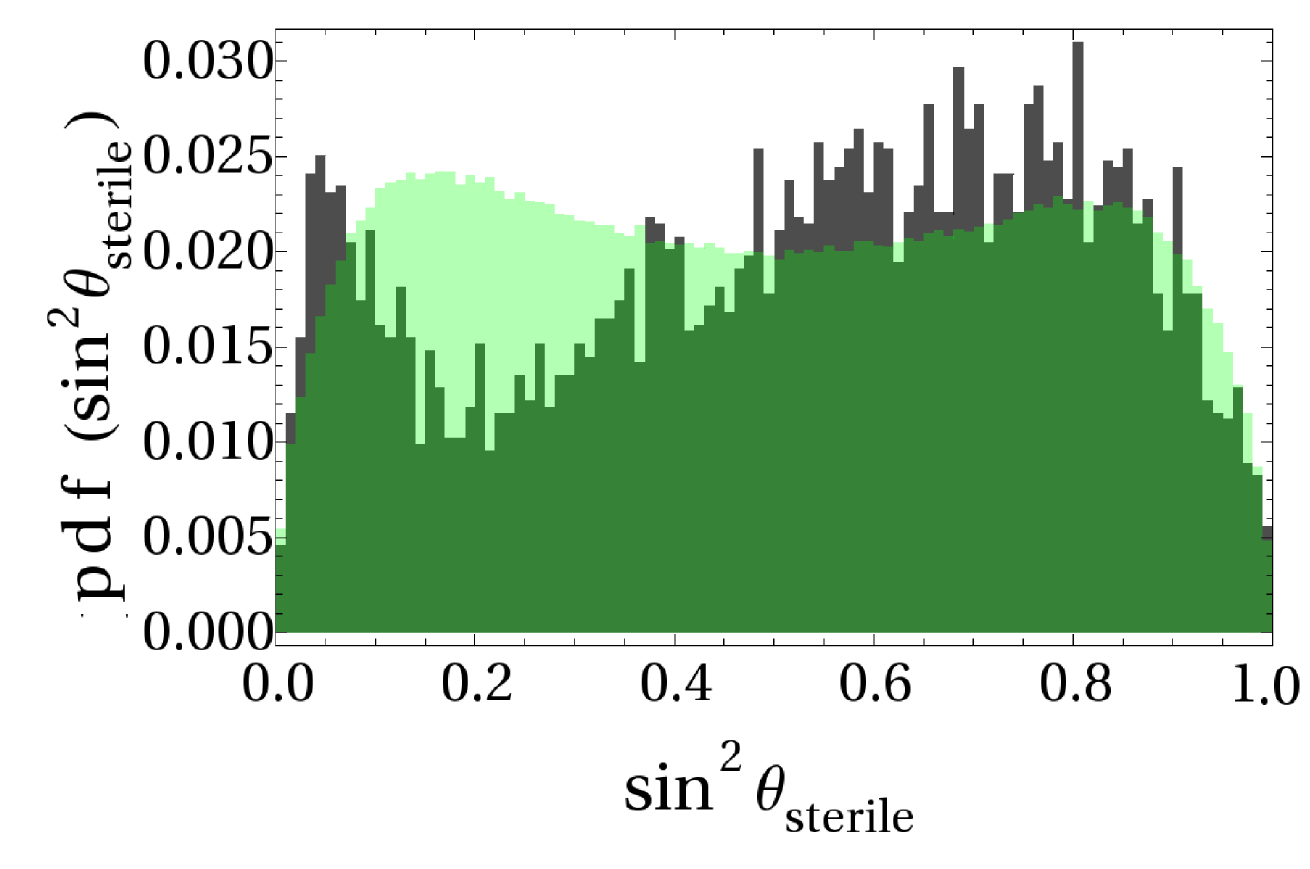}
\caption{Neutrino mass cuts make the sneutrino LSPs more sterile: (i) Run I (ii) Run II. The black plot satisfies the cuts and the light green curve is a random sample that does not satisfy the neutrino constraints.}
\label{pdf}
\end{center}
\end{figure}

\begin{figure}[h]
\begin{center}
\includegraphics[width = 0.9\textwidth]{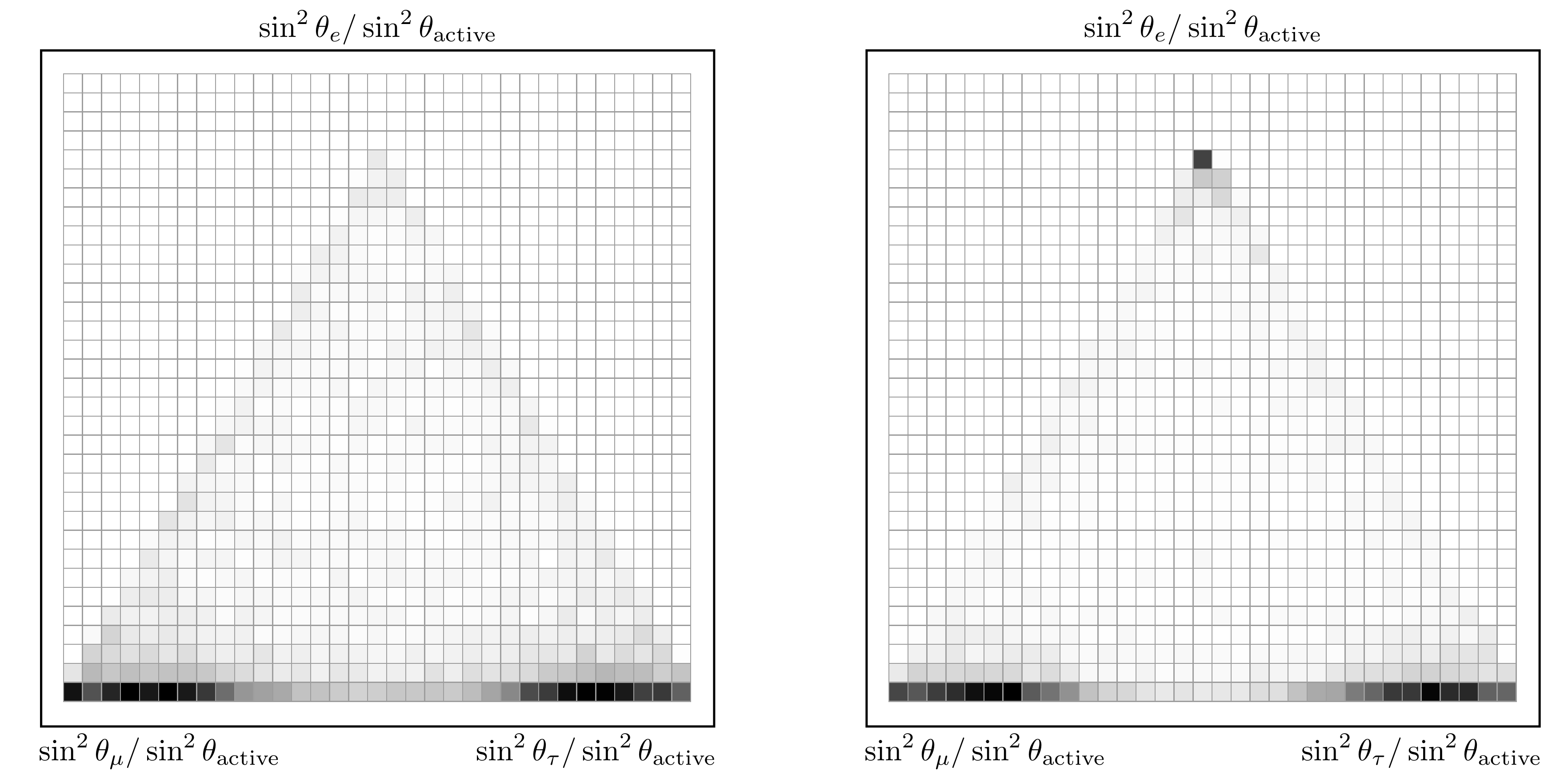}
%\includegraphics[width = 0.45\textwidth]{lapLSP-I.pdf}
%\hspace{0.5cm}
%\includegraphics[width = 0.45\textwidth]{lapLSP-II.pdf}
\caption{$e$-$\mu$-$\tau$ mixing in the active sector for the sneutrino LSPs that satisfy neutrino mass cuts: (i) Run I (ii) Run II}
\label{LSPemt}
\end{center}
\end{figure}

One very interesting feature of our scenario is that the neutrino mass constraints shed light on the active/sterile mixing of the LSP sneutrino. In particular,  the sneutrino LSP tends to be more sterile, as seen in Fig. \ref{pdf}.  This bias toward a mostly sterile LSP is mostly due to the $R$ cut.  Thus, we do find that at least the requirement of a mass hierarchy in the neutrino sector seems to prefer a sterile-dominant LSP, which is in accordance with expectations from relic abundance and Z-width constraints. Furthermore, in the active flavor sector, the sneutrino LSP tends to be predominantly $\mu$ or $\tau$ flavored as seen in Fig. \ref{LSPemt}.

\subsection{Relic abundance}

\begin{figure}
\begin{center}
\includegraphics[width = 0.49\textwidth]{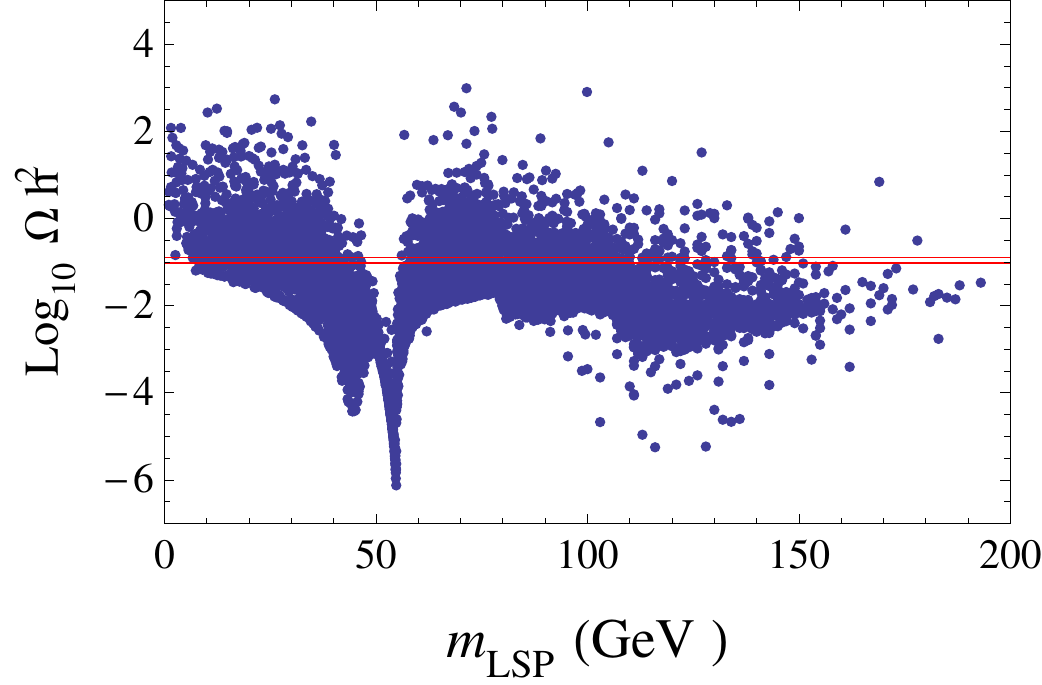}
\includegraphics[width = 0.49\textwidth]{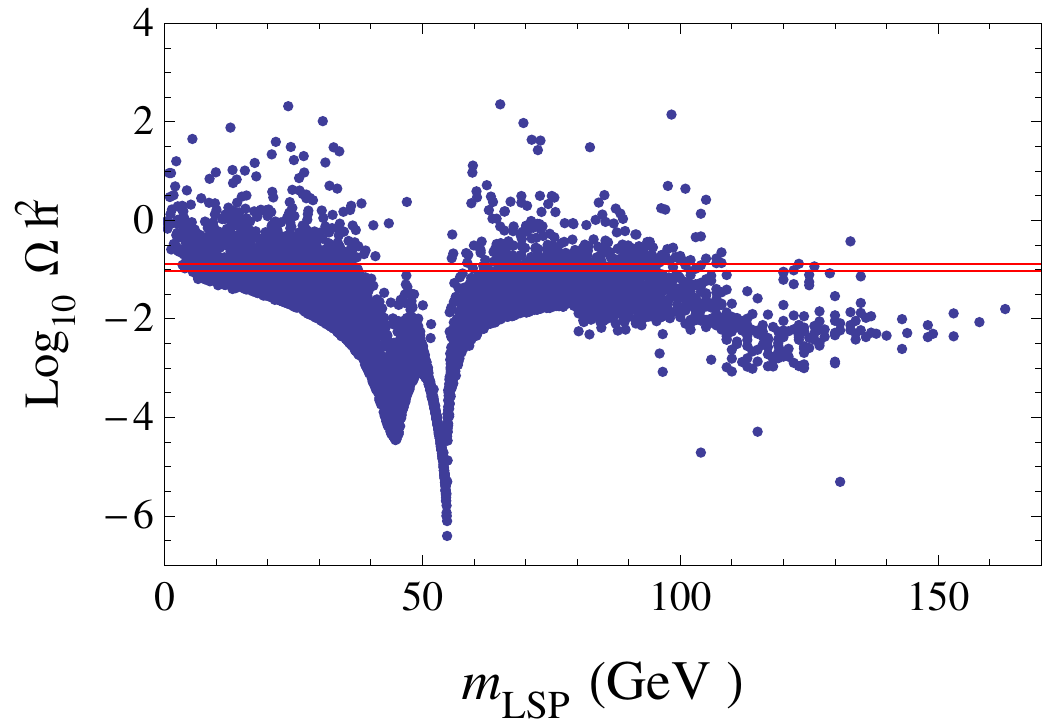}
\caption{Scatter plots of relic abundance values for the sneutrino LSP: (i) Run I (ii) Run II.  The  horizontal red lines indicate the range of $\Omega_{DM} h^{2}$ values allowed by observation. }
\label{DMvsmDM}
\end{center}
\end{figure}

The scatter plots of Fig. \ref{DMvsmDM} show the $\Omega_{DM} h^{2}$ values for the random matrices of Runs I and II.
The funnel shapes at the $Z$ and Higgs poles are as expected. Beyond this, the most striking features of these plots is that the relic abundance constraint selects a very ``ordinary'' region of parameter space. We needn't go into the Higgs or Z poles or out to the edges of the parameter scan to find LSPs with the appropriate relic abundance.

\begin{figure}
\begin{center}
\includegraphics[width = 0.9\textwidth]{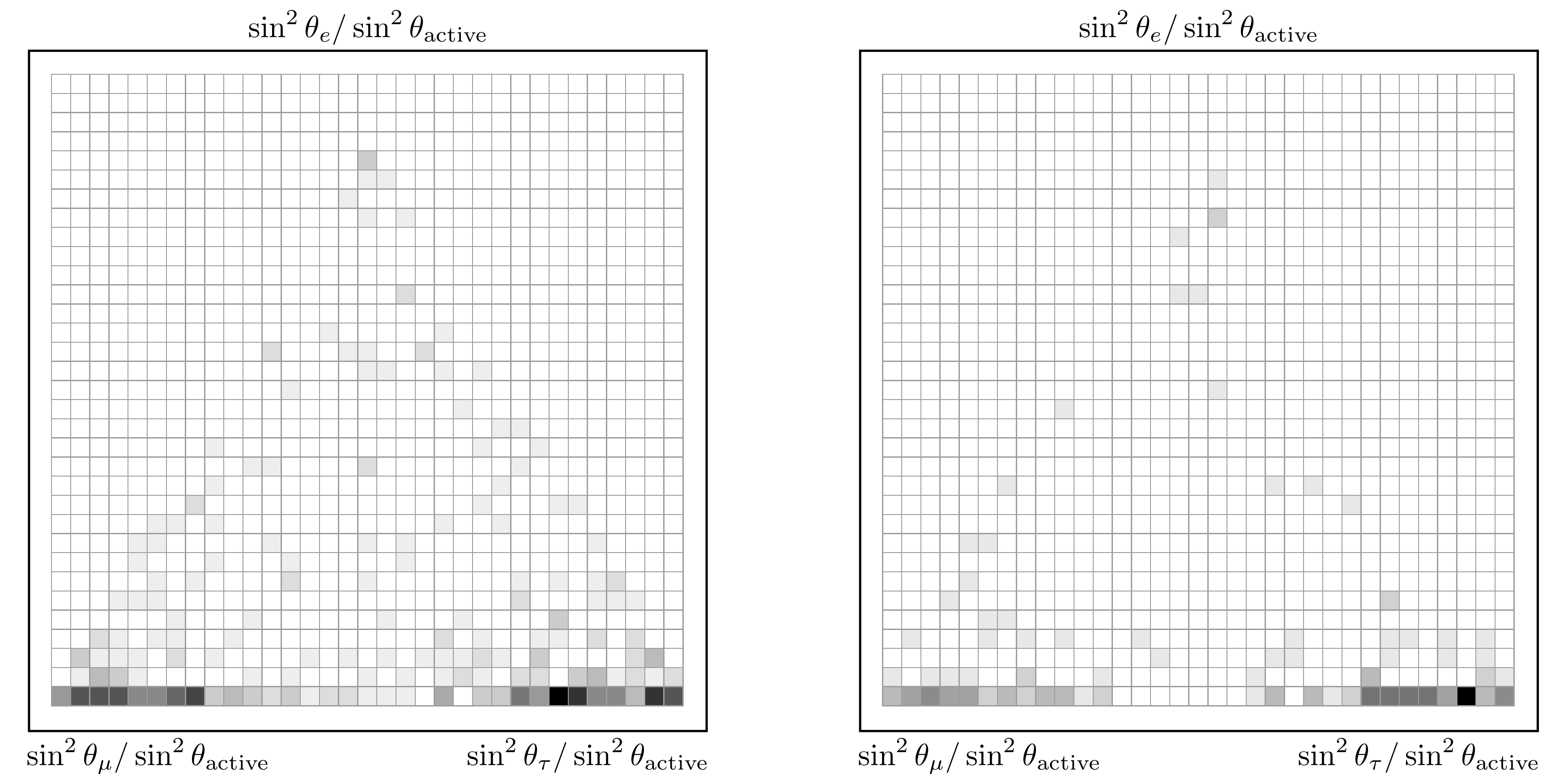}
%\includegraphics[width = 0.45\textwidth]{lap-I.pdf}
%\hspace{0.5cm}
%\includegraphics[width = 0.45\textwidth]{lap-II.pdf}
\caption{$e$-$\mu$-$\tau$ mixing in the active sector for the sneutrino LSPs that satisfy neutrino mass and relic abundance constraints: (i) Run I (ii) Run II}
\label{DMemt}
\end{center}
\end{figure}

As discussed earlier, Table \ref{table:Allstats} shows that a random set of  sneutrino mass matrices is as likely to satisfy the relic abundance constraint as one satisfying the neutrino mass constraints. The $\Omega_{DM} h^{2}$ constraint is furthermore essentially flavor-blind and so does little to
change our expectations for the flavor content of the LSP sneutrino, compared with what  we found after the neutrino cuts alone. This can be seen by comparing  Figures \ref{LSPemt} and \ref{DMemt}.

\begin{figure}
\begin{center}
\includegraphics[width = 0.32\textwidth]{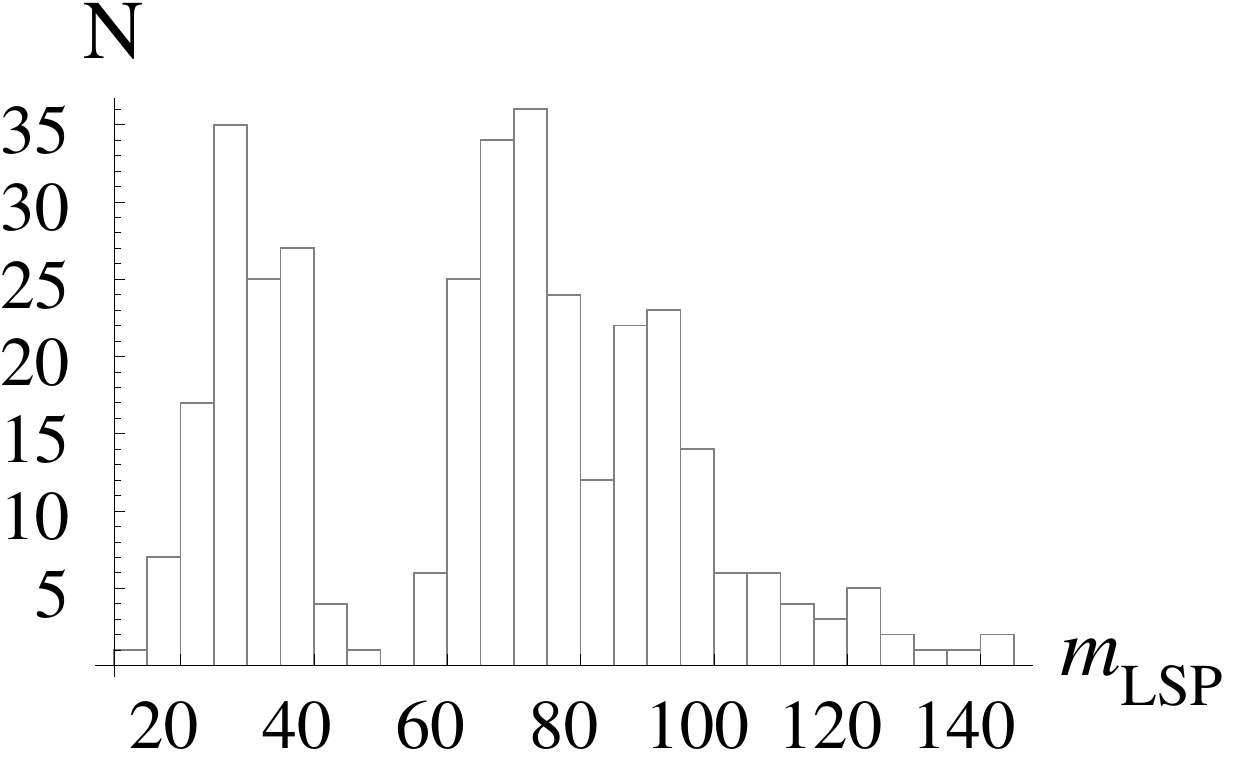}
\includegraphics[width = 0.32\textwidth]{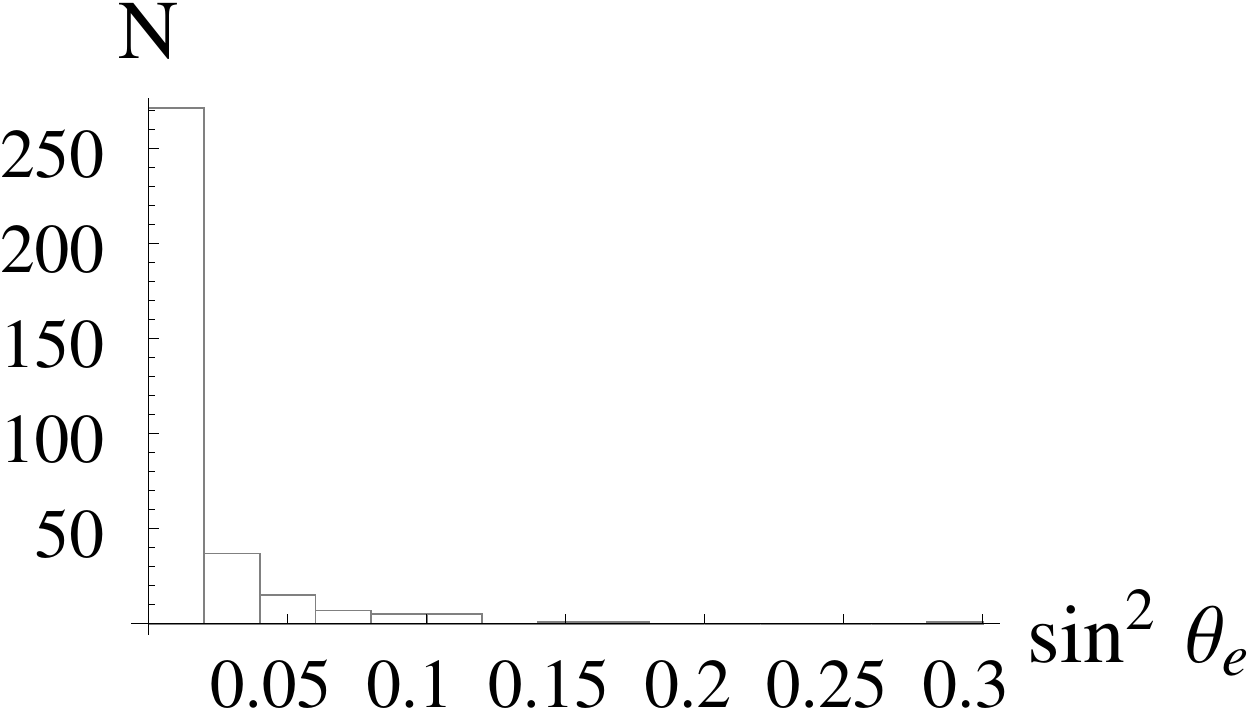}
\includegraphics[width = 0.32\textwidth]{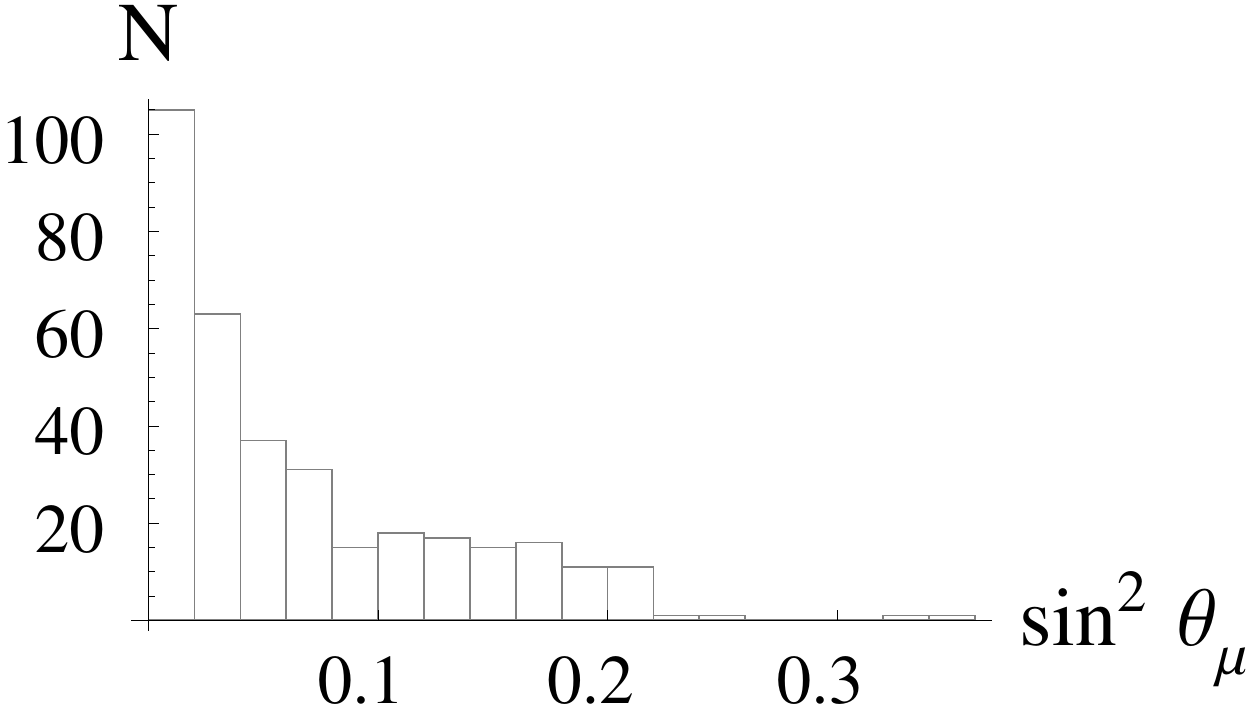}
\includegraphics[width = 0.32\textwidth]{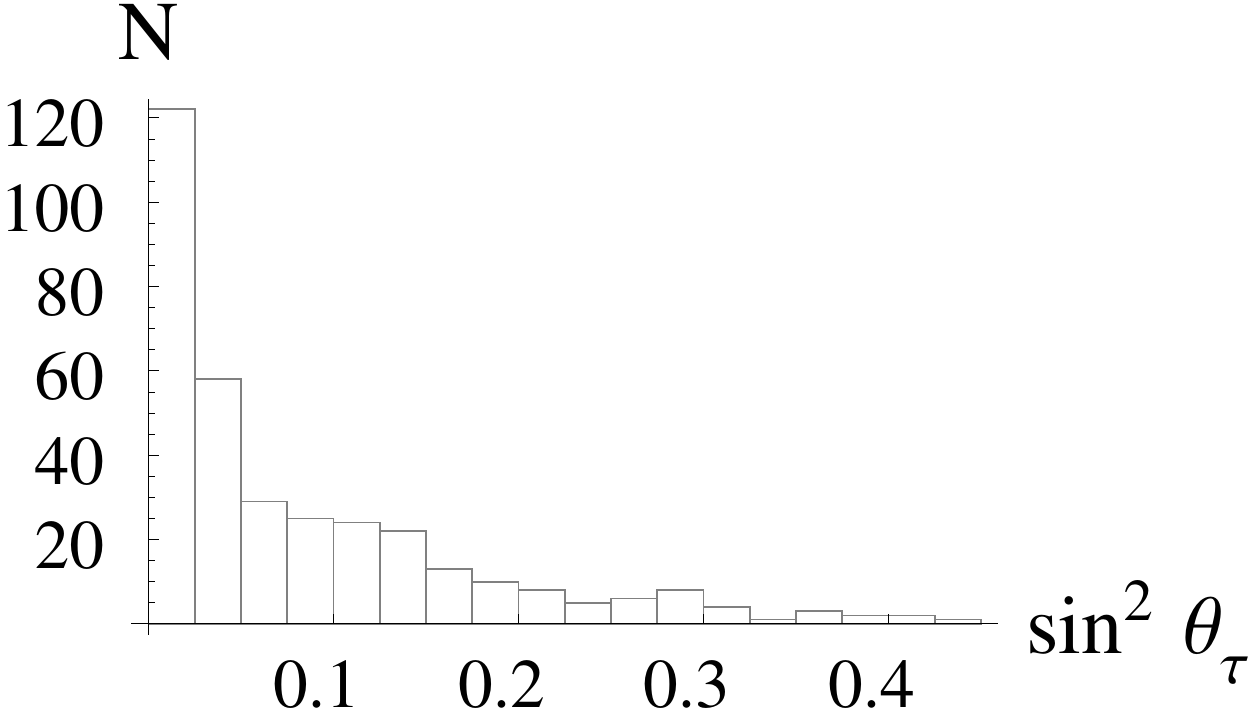}
\includegraphics[width = 0.32\textwidth]{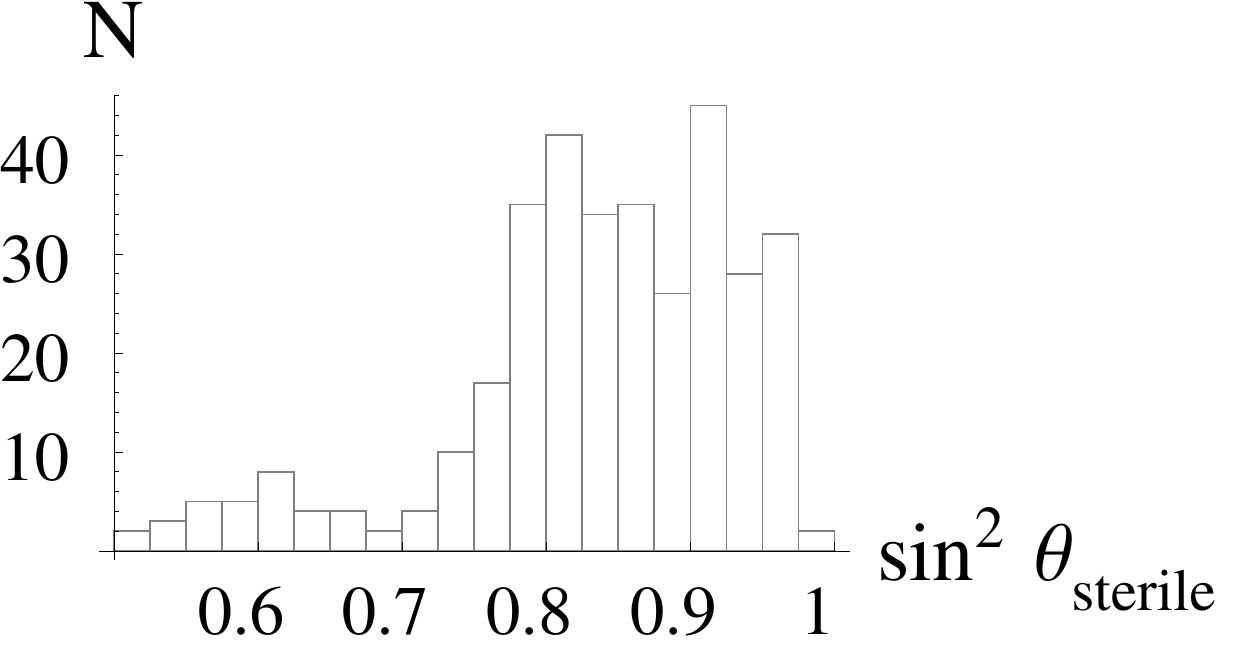}
\includegraphics[width = 0.32\textwidth]{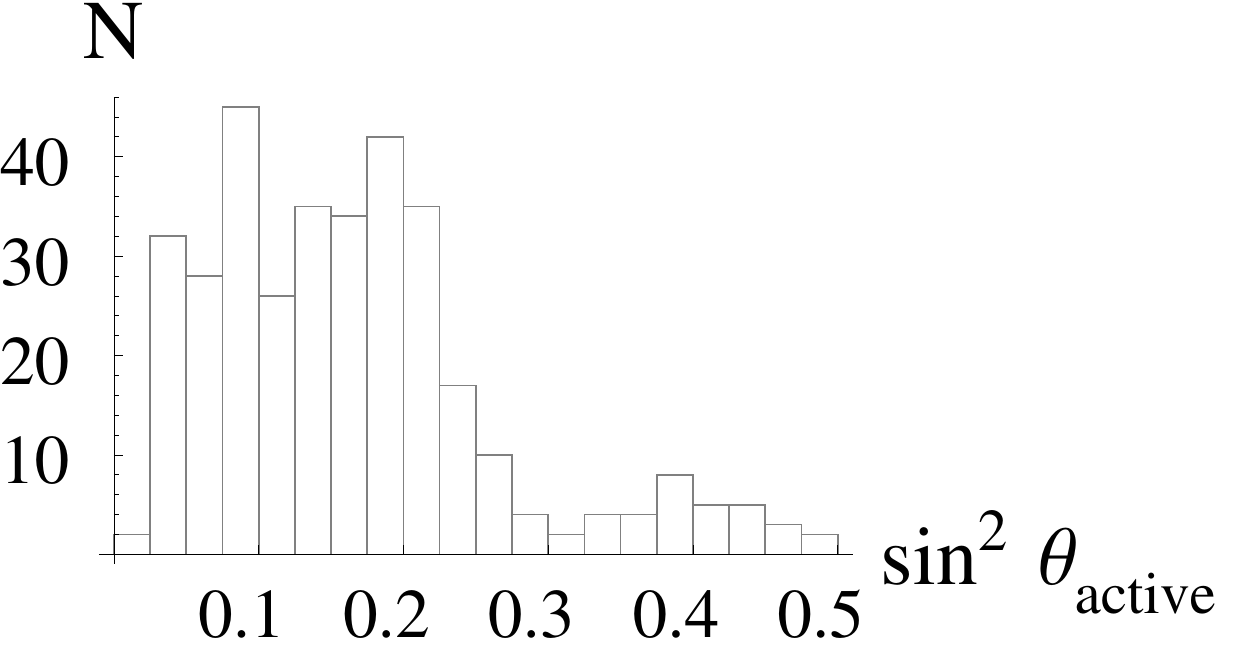}
\caption{Mass and mixing angle distributions for the 343 sneutrino LSPs from the sample that satisfies neutrino mass, relic abundance and invisible $Z$ width cuts in Run I. The LSP mass is in GeV.}
\label{DMhistRunXi}
\end{center}
\end{figure}

\begin{figure}
\begin{center}
\includegraphics[width = 0.32\textwidth]{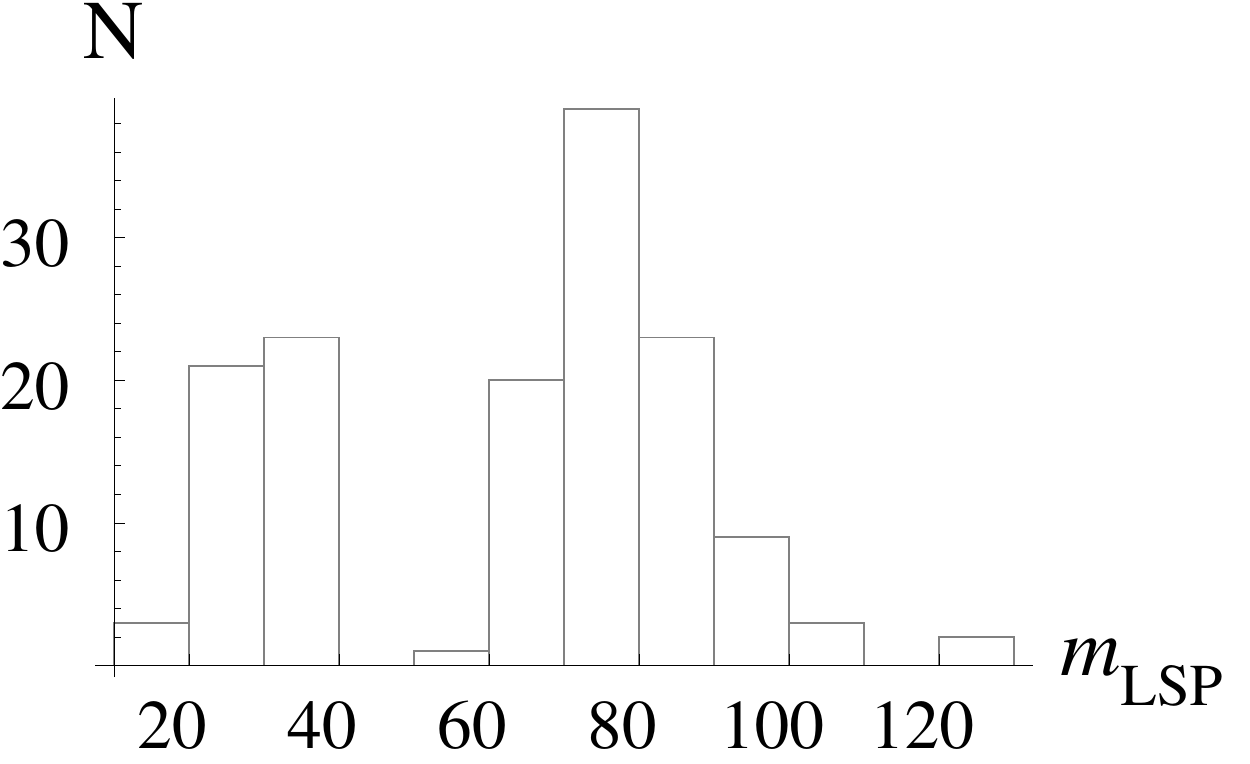}
\includegraphics[width = 0.32\textwidth]{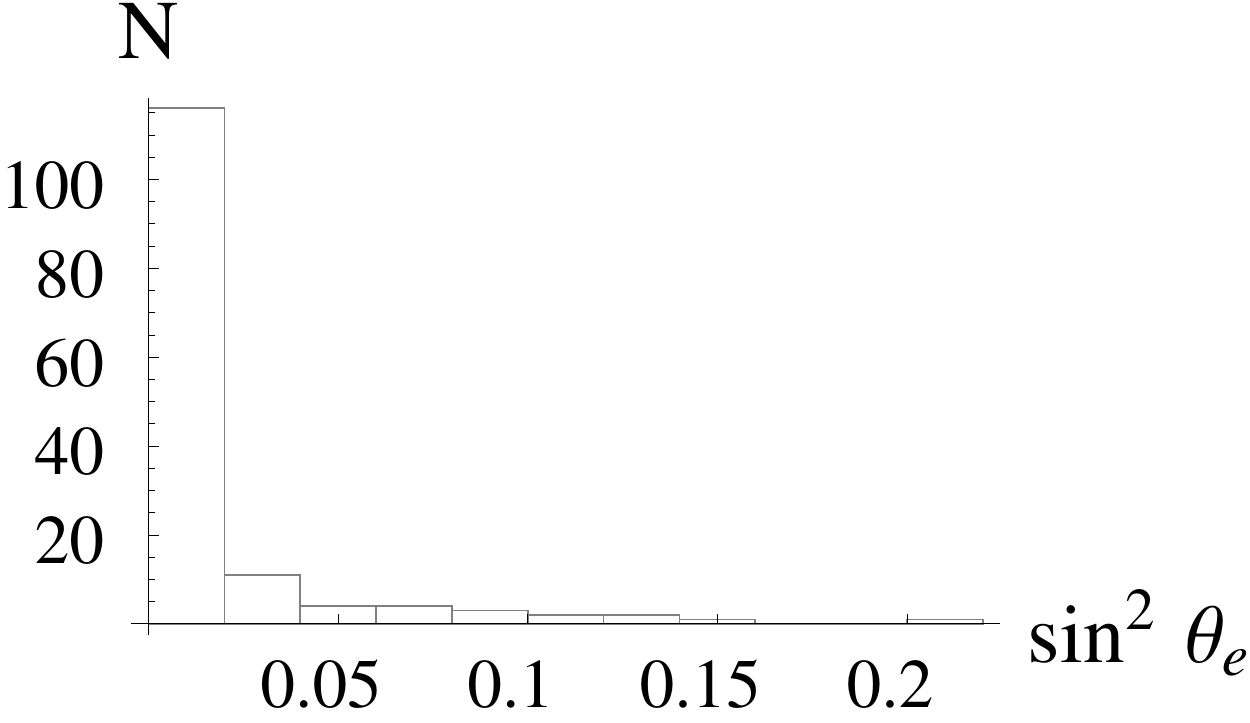}
\includegraphics[width = 0.32\textwidth]{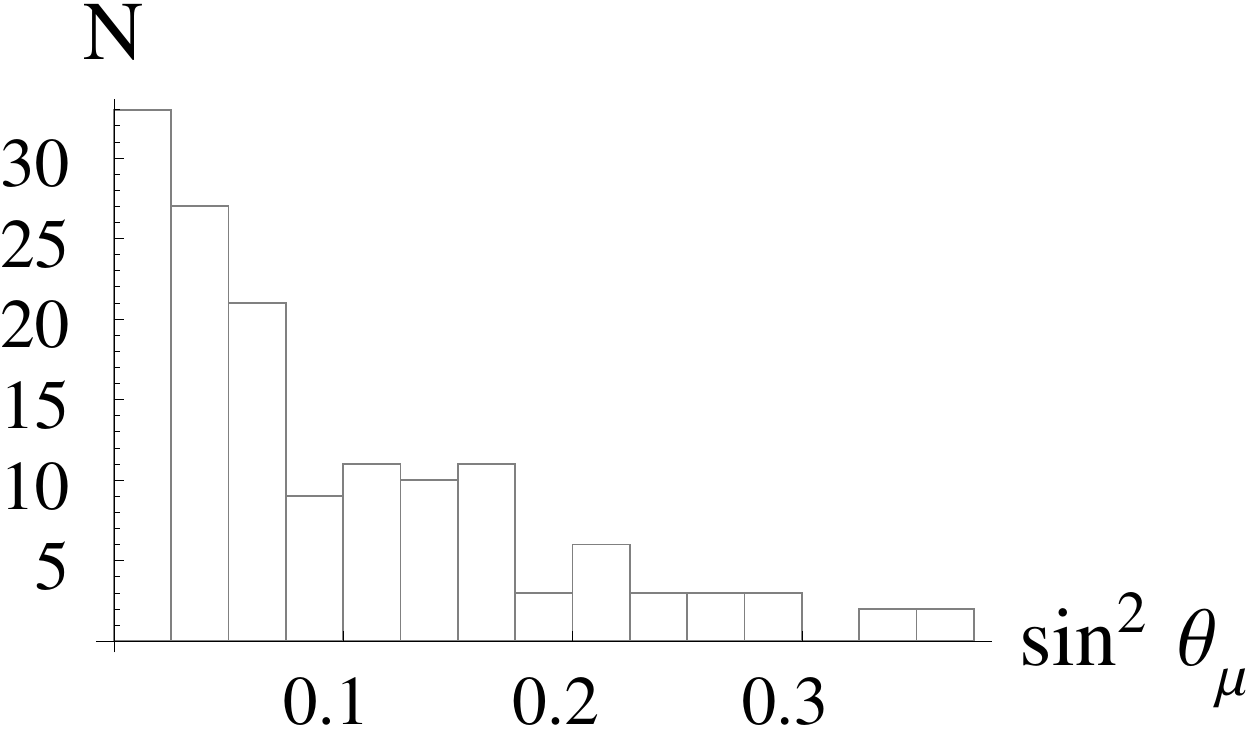}
\includegraphics[width = 0.32\textwidth]{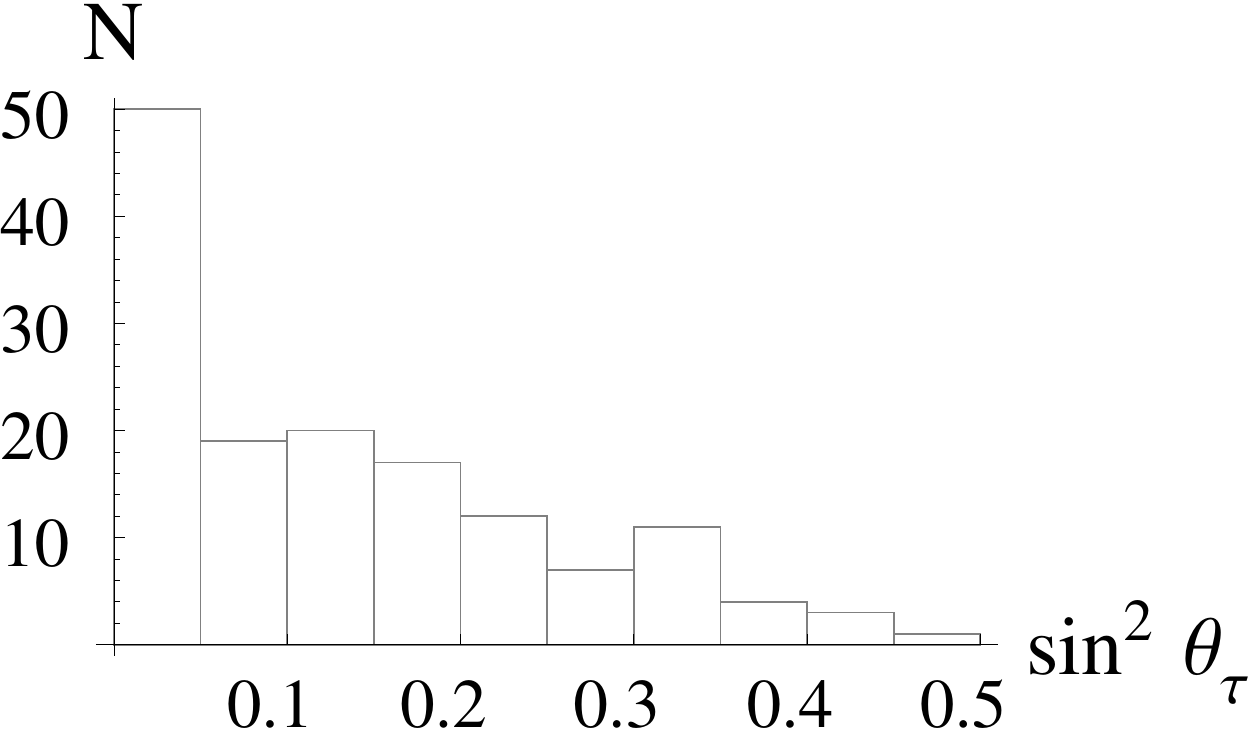}
\includegraphics[width = 0.32\textwidth]{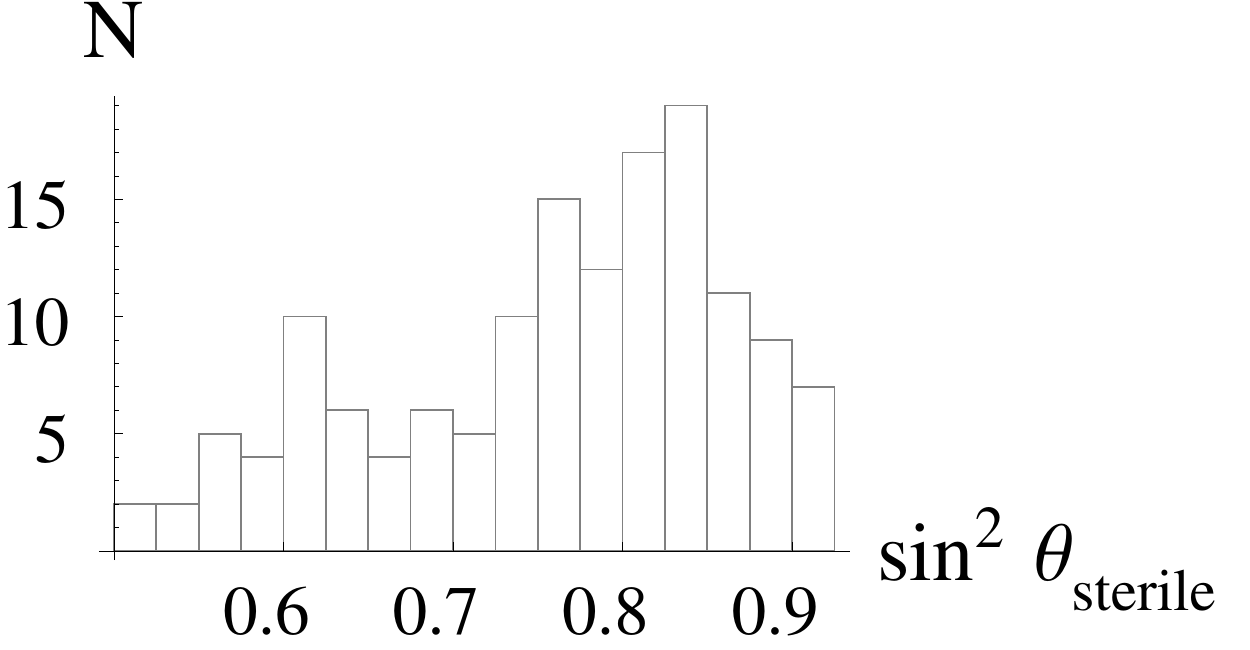}
\includegraphics[width = 0.32\textwidth]{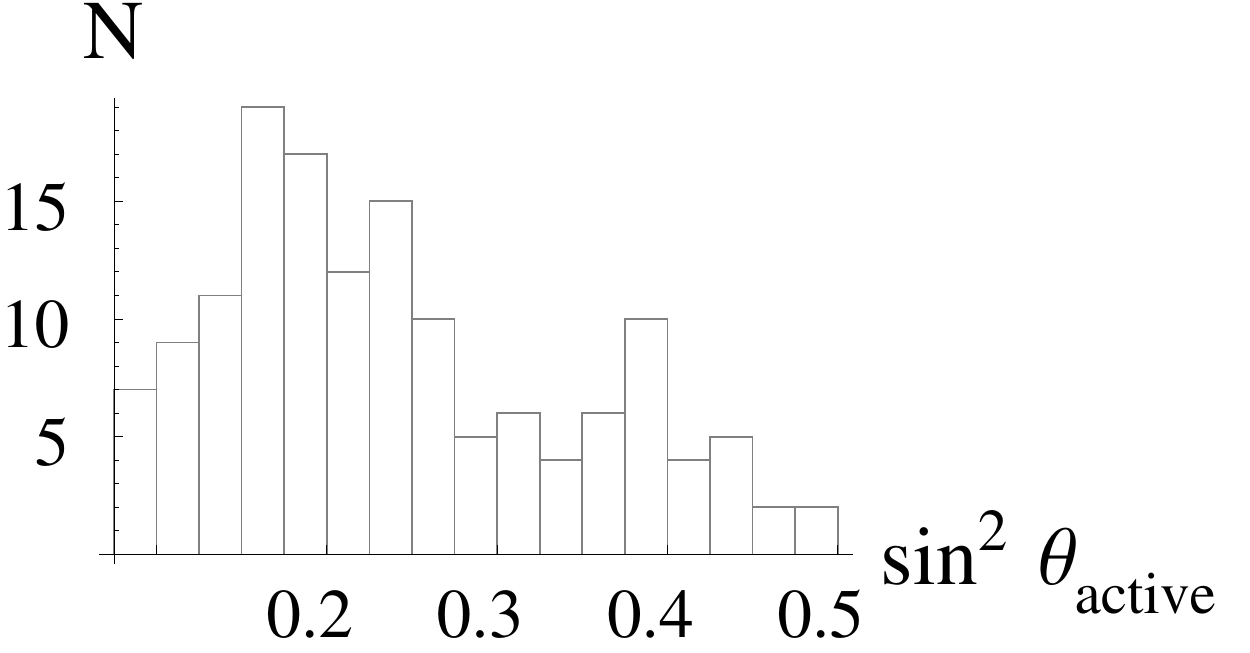}
\caption{Mass and mixing angle distributions for the 144 sneutrino LSPs from the sample that satisfies neutrino mass, relic abundance and invisible $Z$ width cuts in Run II. The LSP mass is in GeV.}
\label{DMhistRunX345}
\end{center}
\end{figure}

The distributions of LSP masses and mixing angles for the sets of mass matrices that satisfy both the neutrino mass and relic abundance constraints are plotted in Figures \ref{DMhistRunXi} and \ref{DMhistRunX345}. The most important result is that the LSP typically has very little electron component, and is dominantly a mixture of $\mu$ and $\tau$. This means that it is likely that cascades to the LSP will be $\mu$ and $\tau$ rich, motivating the collider analyses in  section \ref{sec:collider}.

\begin{figure}
\begin{center}
\includegraphics[width = 0.49\textwidth]{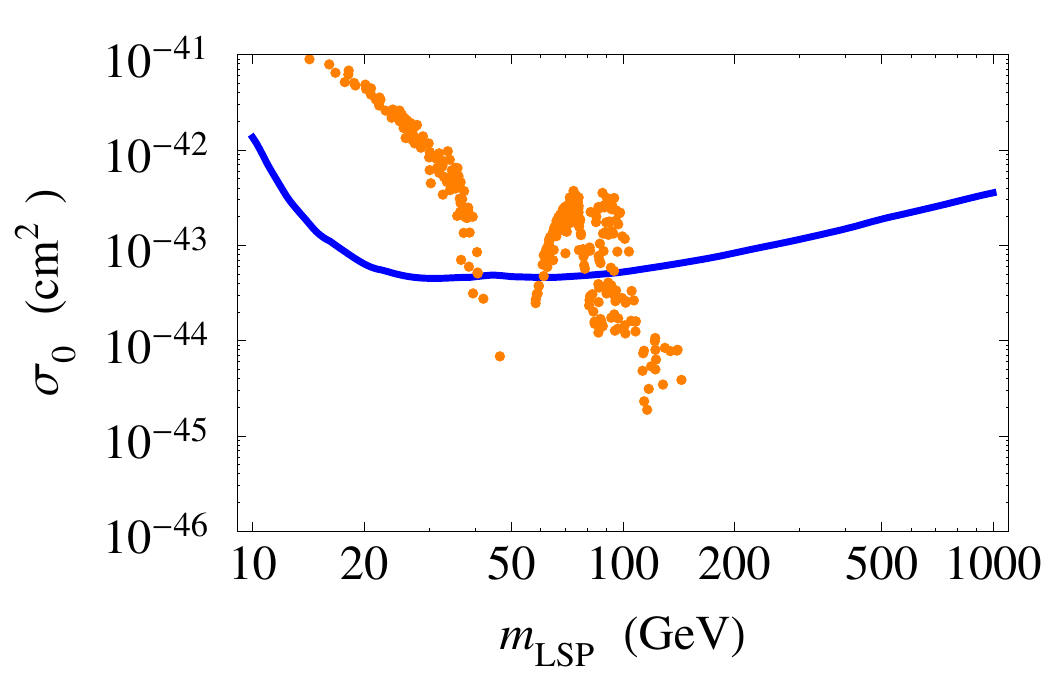}
\includegraphics[width = 0.49\textwidth]{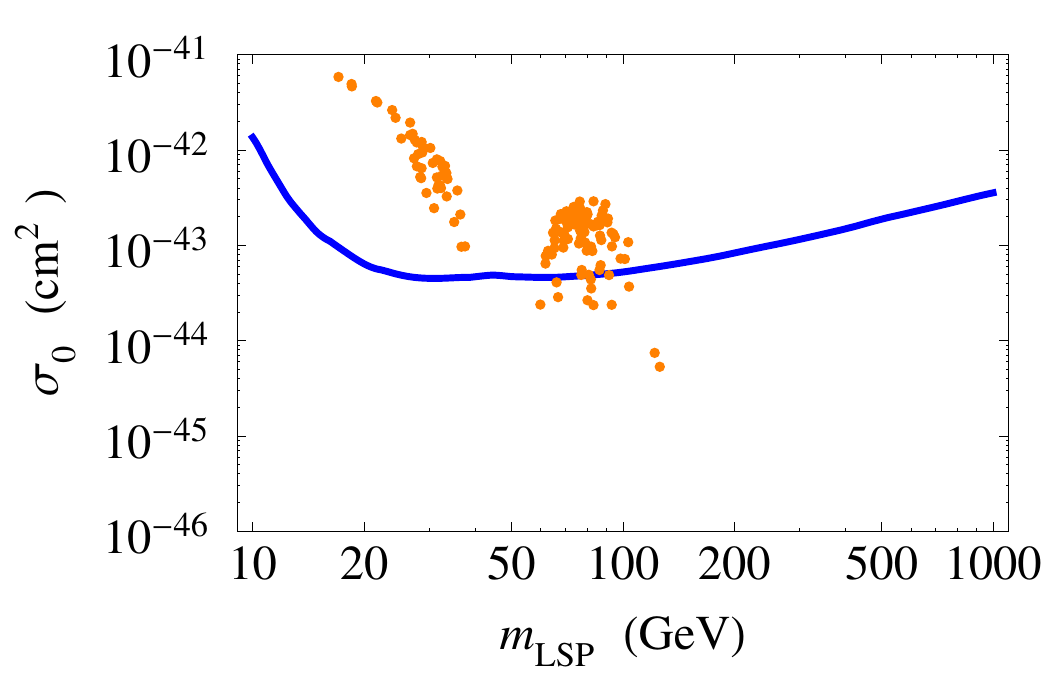}
\caption{Distributions of $\snu1$-nucleon spin-independent cross sections for  Higgs-mediated scattering,  for the samples that satisfy neutrino mass, relic abundance and invisible $Z$ width cuts. The regions above the thick (blue) curve is excluded by CDMS and XENON10. In Runs I (left) and II (right), 69 and 12 LSPs satisfy the direct detection constraints, respectively. }
\label{directDM}
\end{center}
\end{figure}

\begin{figure}
\begin{center}
\includegraphics[width = 0.45\textwidth]{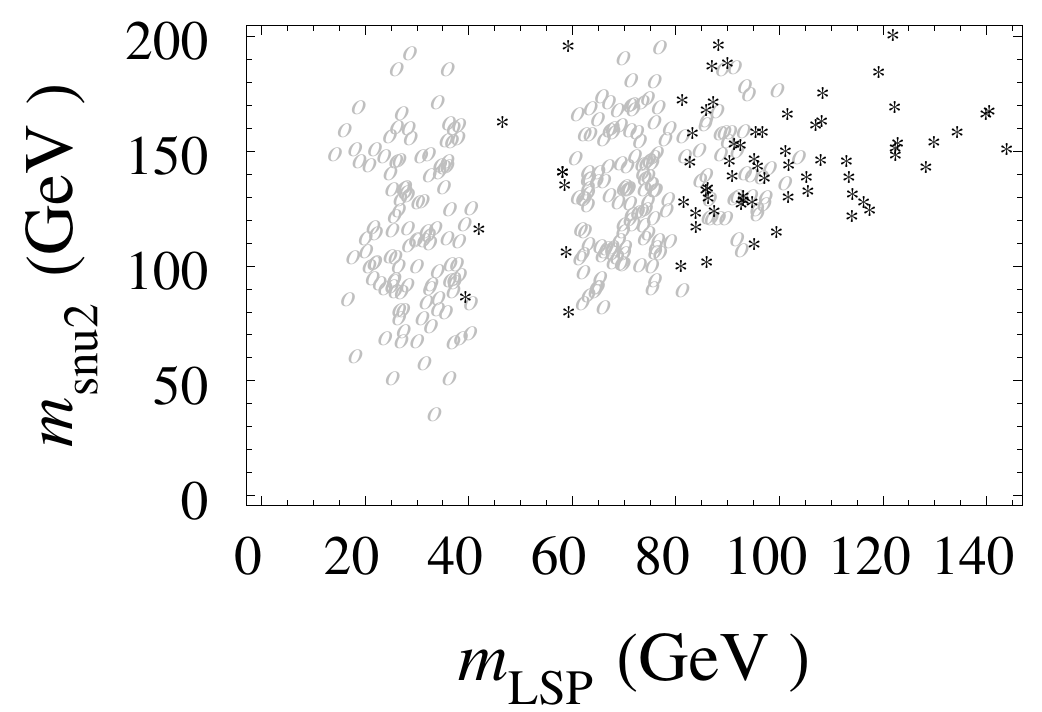}
\hspace{0.5cm}
\includegraphics[width = 0.45\textwidth]{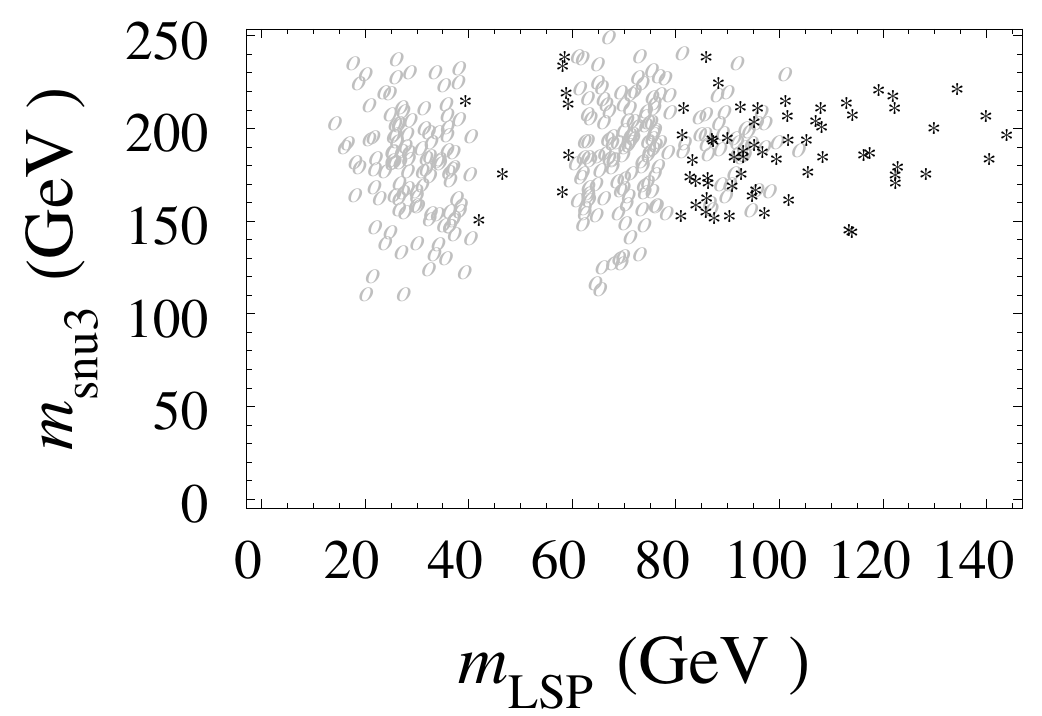}
\includegraphics[width = 0.45\textwidth]{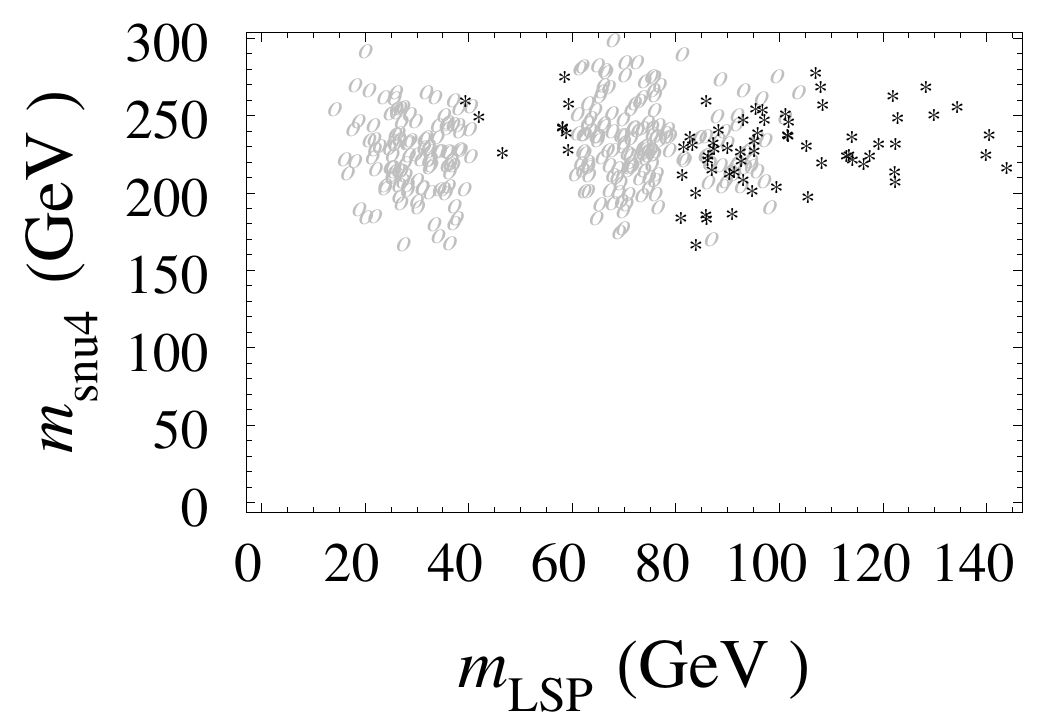}
\hspace{0.5cm}
\includegraphics[width = 0.45\textwidth]{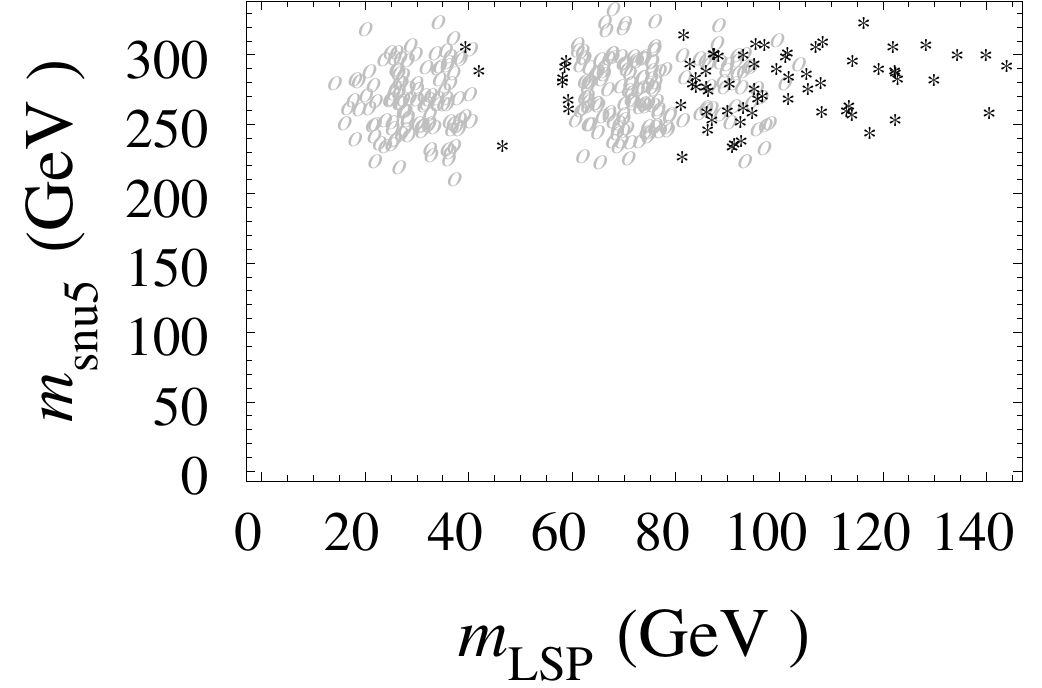}
\includegraphics[width = 0.45\textwidth]{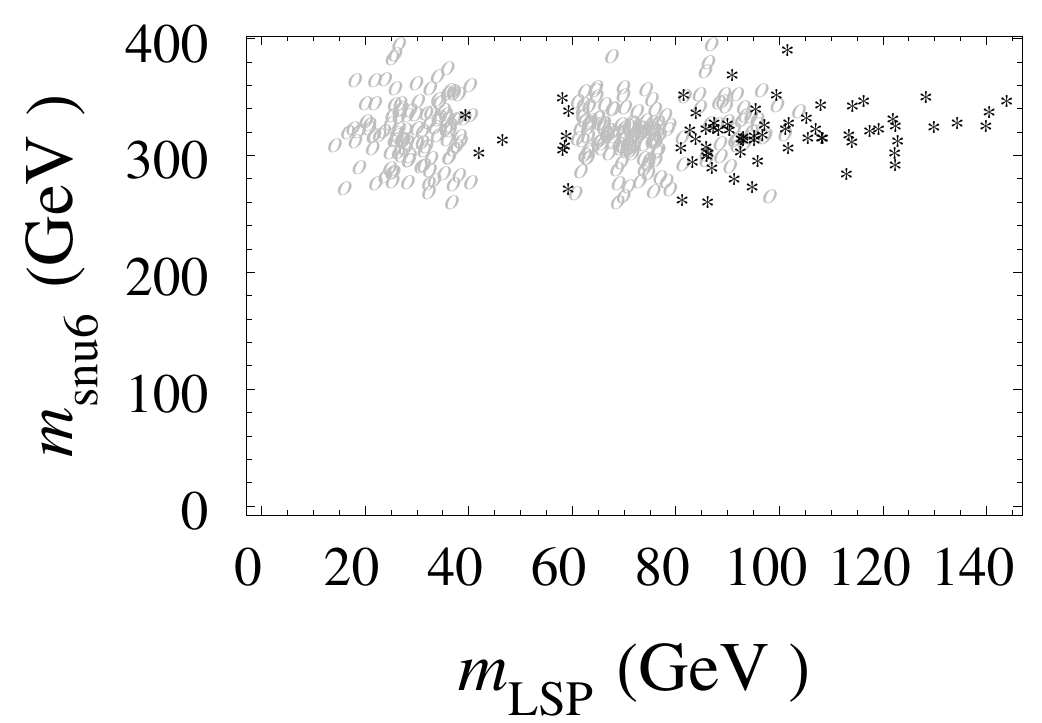}
\caption{Sneutrino mass ratios for the 343 sets of matrices that satisfy neutrino mass, relic abundance and invisible $Z$ width cuts in Run I. The grey circles and black asterisks represent sneutrino LSPs that do not/do satisfy XENON10 and CDMS bounds.}
\label{massratios}
\end{center}
\end{figure}

\begin{figure}
\begin{center}
\includegraphics[width = 0.45\textwidth]{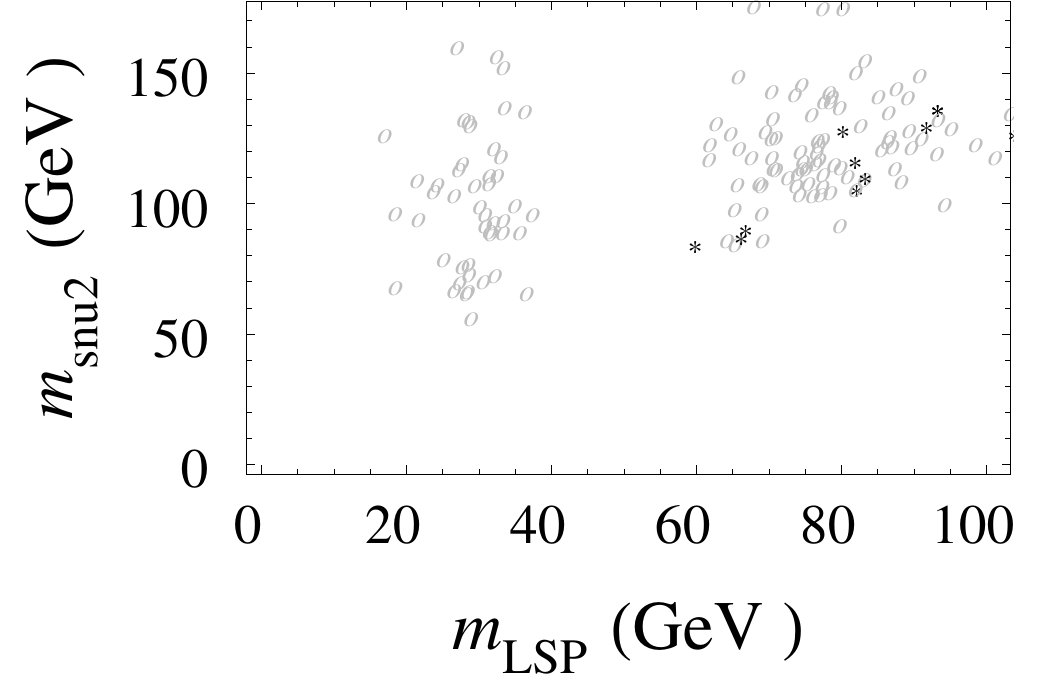}
\hspace{0.5cm}
\includegraphics[width = 0.45\textwidth]{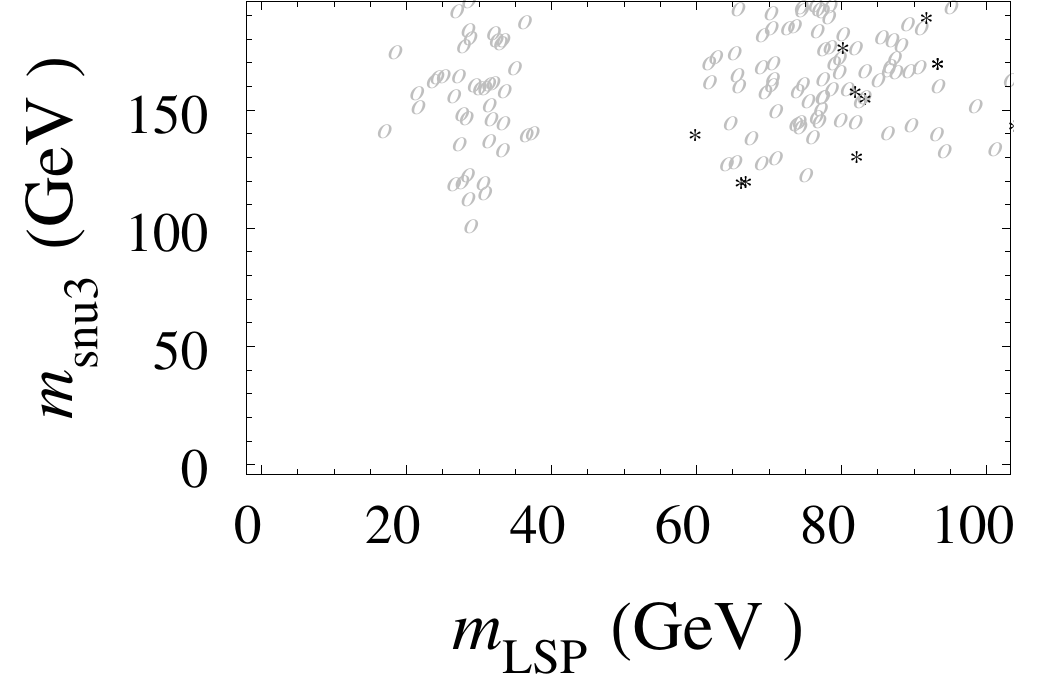}
\includegraphics[width = 0.45\textwidth]{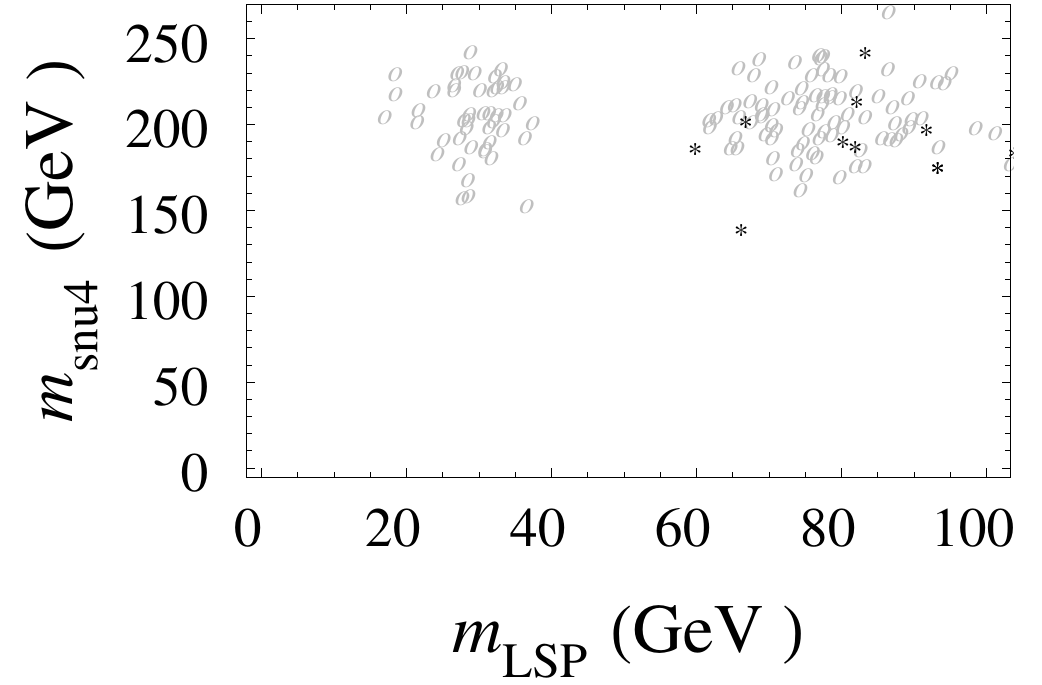}
\hspace{0.5cm}
\includegraphics[width = 0.45\textwidth]{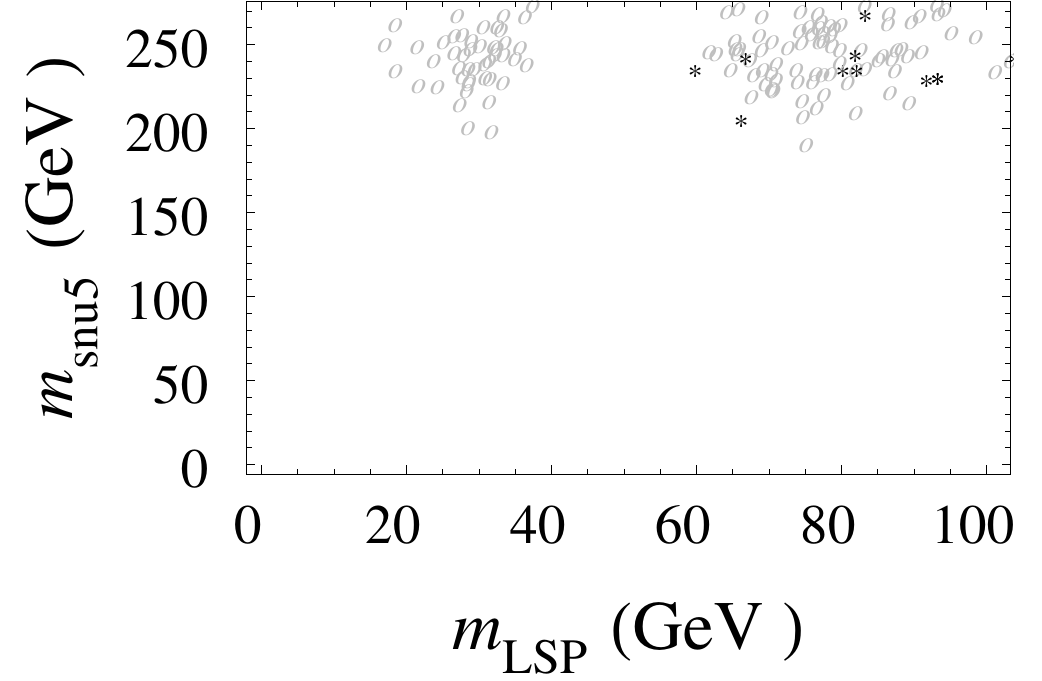}
\includegraphics[width = 0.45\textwidth]{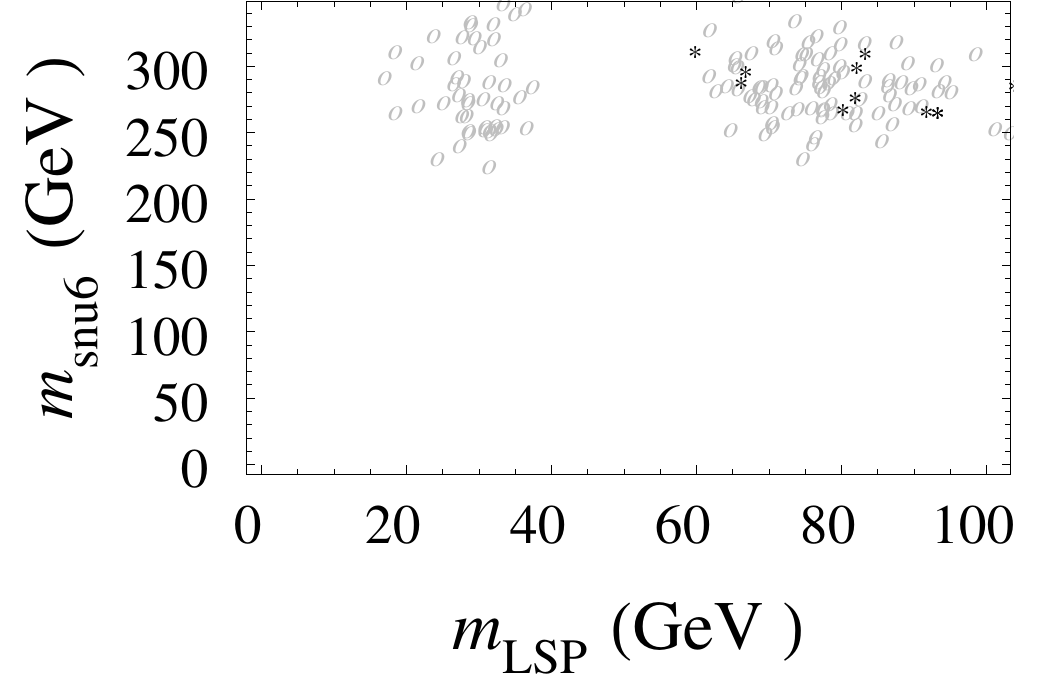}
\caption{Sneutrino mass ratios for the 144 sets of matrices that satisfy neutrino mass, relic abundance and invisible $Z$ width cuts in Run II. The grey circles and black asterisks represent sneutrino LSPs that do not/do satisfy XENON10 and CDMS bounds.}
\label{massratios-RunX345}
\end{center}
\end{figure}

\subsection{Direct detection}

XENON10 \cite{Angle:2007uj} and CDMS \cite{Ahmed:2008eu} put the strongest constraints on spin-independent WIMP-nucleon scattering. Because the $Z$-exchange contribution to the scattering amplitude can be strongly suppressed in the presence of lepton-number violation, we require that the Higgs-mediated contribution satisfy the bounds from these experiments. Fig. \ref{directDM} shows the distributions of LSP-nucleon cross sections for the sets of mass matrices that satisfy the neutrino, relic abundance, and invisible $Z$-width constraints.

Using the same sets of matrices, distributions of sneutrino mass ratios are shown in Figures \ref{massratios} and \ref{massratios-RunX345}.  The vertical gaps in the plots are due to the fact that sneutrino LSPs near the $Z$ or Higgs poles rapidly annihilate to leave a very low relic abundance.
These plots distinguish between the sneutrino LSPs that do and do not satisfy direct detection bounds.
While a significant fraction do in fact satisfy the current direct detection bounds,  the next generation of direct detection experiments will probe down to the $10^{-44} {\rm cm^2}$ level, and can therefore exclude much of the remaining parameter space.

\section{Collider phenomenology}
\label{sec:collider}
The LHC phenomenology of mixed sneutrinos in the MRSSM is potentially very rich. If we expect large flavor mixing in the sneutrino mass matrix,  we should also expect left-handed slepton mass matrices with large off-diagonal entries.  We will see that cascades involving left-handed sleptons can consequently produce opposite-sign dileptons that reveal this lepton flavor violation.

Because experimental constraints on flavor violation are the strongest for the first two generations,  $e$-$\mu$ flavor violation associated with  sleptons produced at the LHC would be a particularly striking signature of this scenario.  Moreover, as we have learned in the previous section, this model tends to give produce an LSP with a significant $\mu$-component, motivating inclusive signatures of this feature. In this section we show that in  certain regions of MRSSM parameter space, these signatures will indeed be observable. Recent work has shown detectability of a $\tau$ flavor violating signal $\cite{Carquin:2008gv}$, and other related work on flavor violation includes $\cite{Kribs:2009zy,Feng:2007ke,Feng:2009bs}$.

\subsection{Slepton mass matrix}

The slepton mass matrix is obtained from the highly mixed sneutrino mass matrix, by enforcing the constant $D$-term mass difference in the left-handed sector. Upon diagonlization, the left-handed mass eigenstates are related to the gauge eigenstates by $\til l_{i} = U_{ie} \til l_{e}$ where $U_{ie}$ diagonalizes the mass matrix, $i=1,2,3.$ Unlike in the MSSM, where the left-handed selectron and smuon are nearly degenerate, the left-handed sleptons have a mass hierarchy that can lead to an observable flavor violating signal.

Although the right-handed slepton mass matrix may also be highly mixed, we take it to be diagonal for simplicity's sake.
The $e$-$\mu$ flavor-violating signal we study below can arise just as easily in the more general case with mixed left- and right-handed slepton mass matrices.

\subsection{Neutralino and chargino masses}

The MRSSM has 4 neutralinos in addition to the MSSM neutralinos. In the absence of a $\mu$ term, the $\til R_{u}^{0}$ and $\til R_{d}^{0}$ pair up with $\til H_{u}^{0}$ and $\til H_{d}^{0}$ respectively, $\mu_{u} \til H_{u}^{0} \til R_{u}^{0}$ and $\mu_{d} \til H_{d}^{0} \til R_{d}^{0}$.
Furthermore, the bino and wino get Dirac masses with two additional neutralino fields.  These Dirac gauginos induce radiative contributions to the sfermion masses-squared that are a factor $\sim g^2/(16 \pi^2)$ smaller than the gaugino masses-squared themselves.
The charginos are similarly paired with the charged components of the additional fields.

\subsection{$e$-$\mu$ flavor violation}

\begin{figure}[h]
\begin{center}
\includegraphics[width = 0.50\textwidth,viewport=70pt 385pt 510pt 780pt]{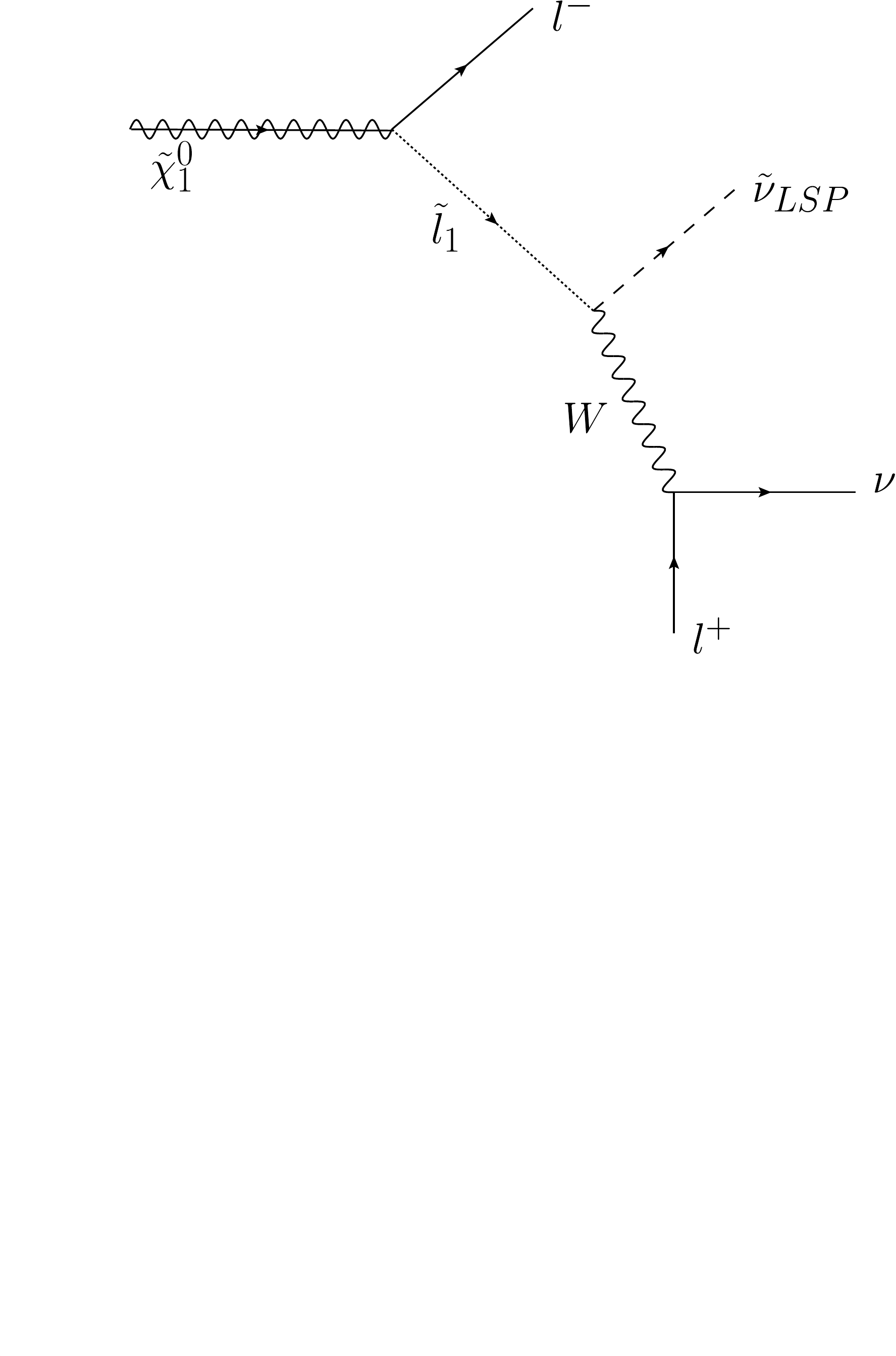}
\caption{A decay chain with potential lepton flavor violation}
\label{LFVdecay}
\end{center}
\end{figure}

A decay chain involving left-handed sleptons and ending in $\snu1$ can provide a signal of lepton flavor violation.
An important example for the sample parameter point we study below is the chain $\til \chi_{1}^{0} \rightarrow \til l_{1} l \rightarrow \snu1 \nu l,$ where $\til \chi_{1}$ is the lightest neutralino and $\til l_{1}$ is the lightest left-handed slepton (see figure \ref{LFVdecay}).
As the $W$-boson in this decay chain has flavor-universal couplings, flavor violation occurs at only the first vertex, $\til \chi_{1}^{0}-\til l_{1}-l$.  The sneutrino vertex itself does not contribute to the flavor-violating effect. The  signal we will focus on therefore provides information on the flavor structure of the lightest left-handed slepton, $\til l_{1}$.

Having a larger wino than bino component in the lightest neutralino/chargino increases the coupling to the left-handed sleptons and favors more flavor violating events, given that we are taking the right-handed slepton mass matrix to be flavor-diagonal.
The $e$-$\mu$ flavor-violating signal is also enhanced if the lightest neutralino  couples to the leptons dominantly via gauge couplings rather than Yukawa couplings, so that  decays to  $\tau$'s are suppressed. So, it is advantageous if $\til \chi_{1}^{0}$ has  a larger $\til H_{u}^{0}$ than $\til H_{d}^{0}$ component.
These features are realized in the parameter point that we adopt for our detailed analysis.

\subsection{Particle spectrum}

We now describe the most important details of our sample parameter point. The parameters that determine the neutralino and chargino properties are:
\begin{center}
{\centering $\tan \beta = 10$ $\qquad \mu_{u}= 348$ GeV  $\qquad\mu_{d} = 359$ GeV}\\
{\centering $mD_{wino} = 1$ TeV $\hspace{10.5mm} mD_{bino}= 2$ TeV}
\end{center}
\noindent where $mD_{wino}$ and $mD_{bino}$ are the Dirac masses of the wino and bino.

The bino is heavier than the wino in this setup, but this is not at odds with gauge unification. The new fields in Dirac gaugino models pair up with additional ``bachelor" fields to form complete multiplets and perturbative unification is possible for gauge groups $SU(3)^{3}$ $\cite{Fox:2002bu}$ or $SU(5).$ The relative sizes of the gaugino masses are not entirely determined and depend on the interactions with the ``bachelor" fields. Thus a bino heavier than the wino is still consistent with unification of gauge couplings.

We adopt one of the randomly generated sets of sneutrino mass matrices that passes the relic abundance, neutrino mass, $Z$-width and direct detection cuts described in Section \ref{sec:results}. The left-handed slepton mass matrix is then fixed accordingly. The sparticle masses and the flavor structure of the LSP sneutrino are given in Table \ref{table:spectrum}. In the parameter point considered, the lightest neutralinos and charginos are predominantly of Higgsino-like and their masses are much lower than the TeV scale. However, these light masses are consistent with the anarchic sfermion mass matrices since
flavor observables in the MRSSM are suppressed by the gaugino masses, which are $\mathcal O$($\tev$).

We include only the lightest four neutralinos, the lightest two charginos, and the lightest three sneutrinos in our analysis as the heavier ones are not produced appreciably. Truncated in this way,  the spectrum is  easy to compare with an MSSM spectrum. Note that the neutralinos and charginos have almost degenerate masses and {\em e.g.} decays from a heavier to a lighter neutralino are not allowed kinematically.  This feature is not essential for the $e$-$\mu$ flavor-violating signal, however.

The masses of the lightest neutralinos and charginos are approximately set by the values of $\mu_{u}$ and $\mu_{d}$. These parameters are constrained by electroweak symmetry breaking and cannot be too large. Increasing $\mu_{u}$ relative to $\mu_{d}$ increases the number of $\tau$ decays due to the Yukawa coupling of $\til H_{d}^{0}$, the lightest neutralino. Increasing the masses of the lightest neutralinos and charginos opens up new decay channels through the other mixed sleptons (rather that just through $\til l_{1}$), resulting in a richer LFV signal with multiple sources of flavor violation.

\begin{table}
\centering
\begin{minipage}{0.45\textwidth}
\centering
\begin{tabular}{|c|c|}
\hline
Sparticle & Mass (GeV) \\ \hline
$ m_{\til qL,R}$  &  $1\;\tev$  \\
$ m_{\til g} $ & $1\;\tev$ \\
\hline
$m_{\neut12}$ & 347 \\
$m_{\neut34}$ & 359 \\
\hline
$ m_{\charg1}$ & 346 \\
$ m_{\charg2}$ & 359 \\
\hline
$ m_{\til lR1,2,3}$ & 253 \\
\hline
$ m_{\til l1}$ & 186 \\
$ m_{\til l2}$ & 311 \\
$ m_{\til l3}$ & 375 \\
\hline
$ m_{\snu1}$ & 78  \\
$ m_{\snu2}$ & 124 \\
$ m_{\snu3}$ & 227 \\
\hline
\end{tabular}
\end{minipage}
\begin{minipage}{0.45\textwidth}
\centering
\begin{tabular}{|c|c|}
\hline
\multicolumn{2}{|c|}{$\tilde \nu_{1}$ flavor structure} \\
\hline
$\rm sin^{2}\theta_{\rm active}$ & 0.11 \\
$\rm sin^{2}\theta_{\rm sterile}$ & 0.89 \\
\hline
\hline
$U_{1e}$ & -0.05 \\
$U_{1\mu}$ & -0.324 \\
$U_{1\tau}$ & 0.084 \\
\hline
\end{tabular}
\end{minipage}
\caption{Superpartner masses and sneutrino LSP flavor structure for the chosen parameter point. $U$ relates the gauge and mass eigenstates.}
\label{table:spectrum}
\end{table}

For this parameter point the LSP $\snu1$ is primarily sterile with a large $\mu$ component in the active flavor sector.   More important for the collider phenomenology is the fact that the lightest left-handed slepton also has a large $\mu$ component and tiny $e$ component.  Decays to left-handed sleptons  primarily go through $\til l_{1}$ since it is the lightest, so we should expect an excess of muons over electrons.  On the other hand, decays to right-handed sleptons  proceed via $\til \tau_{2} (= \til \tau_{R} = \til l_{R3})$ due to the large Yukawa coupling.

The mixing angles for the MSSM components of the lightest four neutralinos are given in Table \ref{table:mixing}. The lightest neutralino $\neut12$ is mainly $\til H_{u}$ and has a larger wino than bino component. The wino component can be increased by increasing the Dirac bino mass. As already mentioned, a larger wino component increases the  coupling of neutralinos to left-handed sleptons to produce a stronger LFV signal.

\begin{table}
\centering
\begin{tabular}{|c|c|c|c|c|}
\hline
Neutralino & $\til B^{0}$ & $\til W^{0}$ & $\til H^{0}_{d}$ & $\til H^{0}_{u}$ \\
\hline
$\neut12$ & $\mp$ 0.00275 & $\pm$ 0.02212 & 0.0079 & -0.7041 \\
\hline
$\neut34$ & $\pm$ 0.00025 & $\mp$ 0.00204 & -0.707 & -0.0085 \\
\hline
\end{tabular}
\label{table:mixing}
\caption{The composition of the four lightest neutralinos for the chosen parameter point. The mass eigenstates also contain mixtures of the Dirac partners of $\til B^{0}$, $\til W^{0}$, $\til H^{0}_{d}$, and $\til H^{0}_{u}$.}
\end{table}

\subsection{Branching Ratios}

The relevant branching ratios for the chosen parameter point are given in Table \ref{table:br}. We will be particularly interested in flavor violation in events with  $ee$, $\mu \mu$, and $e \mu$ pairs, which arises dominantly from the $\neut12 $ decay chain depicted in Figure \ref{LFVdecay}.

\begin{table}[h]
\centering
\begin{minipage}[c]{0.4\textwidth}
\centering
\begin{tabular}{|c|}
\hline
$\bf{\neut12 \ra \til l_{1}, \til \tau_{2}}$\\
\hline
$\begin{aligned}
& \neut12 \xra{0.246/0.14} \til l_{1} (\mu/\tau) \xra{0.66} W \snu1 \nonumber \\
  & \qquad\qquad\qquad\qquad\hspace{0.3cm} \xra{0.11}  \snu1 l \nu  \nonumber \\
& \neut12 \xra{0.005} \til \tau_{2} \tau \xra{0.495} \snu2  \tau \nu  \nonumber \\
  & \qquad\qquad\qquad \xra{0.425} \til l_{1} \tau \tau  \xra{\til l_{1} \rm decays} ... \nonumber \\

\end{aligned}$\\
\hline
\end{tabular}
\end{minipage}
\hspace{1cm}
\begin{minipage}[c]{0.4\textwidth}
\centering
\begin{tabular}{|c|}
\hline
$\bf{\neut34 \ra \til l_{1}, \til\tau_{2}}$\\
\hline
$\begin{aligned}
& \neut34 \xra{10^{-4}/0.423} \til l_{1} (\mu/\tau) \xra{0.66} W \snu1 \nonumber \\
  & \qquad\qquad\qquad\qquad\hspace{0.6cm} \xra{0.11} \snu1 l \nu  \nonumber \\
& \neut34 \xra{0.56} \til \tau_{2} \tau \xra{0.495} \snu2  \tau \nu  \nonumber \\
  & \qquad\qquad\qquad \xra{0.425} \til l_{1} \tau \tau  \xra{\til l_{1} \rm decays} ... \nonumber \\
\end{aligned}$\\
\hline
\end{tabular}
\end{minipage}\\

\vspace{0.5cm}
\centering
\begin{tabular}{|c|}
\hline
$\bf{\charg1 \rm \bf{decays}}$\\
\hline
$\begin{aligned}
& \charg1 \xra{0.935} \til l_{1}\nu \nonumber \\
\end{aligned}$\\
\hline
\end{tabular}

\vspace{0.5cm}
\centering
\begin{tabular}{|l|l|}
\hline
\multicolumn{2}{|c|}{$\bf{\til\chi_{2}^{-} \rm \bf{decays}}$} \\
\hline
$\til\chi_{2}^{-}  \xra{0.56} \til \tau_{2} \nu \xra{\til \tau_{2} \rm decays} ... $ & $\til\chi_{2}^{-}  \xra{1.8\times 10^{-4}/0.014} \snu1 (\mu/\tau)$ \\
$\til\chi_{2}^{-}  \xra{3.6\times 10^{-4}/0.215} \snu2 (\mu/\tau) \xra{0.1} \snu1 l^{+} l^{-} $ & $\til\chi_{2}^{-}  \xra{1.8\times 10^{-4}/0.215} \snu3 (\mu/\tau) \xra{0.036} \snu1 l^{+} l^{-}$ \\
\hline
\end{tabular}
\caption{Branching ratios for the neutralino and chargino show flavour violation at the left-handed slepton, $\til l_{1}$, vertex}
\label{table:br}
\end{table}

As shown in Table \ref{table:br}, $\neut12$ decays to $\til l_{1}\mu$ and  $\til l_{1}\tau$ with about the same branching ratios.
In the absence of Yukawa couplings, the branching ratio to $\til l_{1}\mu$ would be much larger because $\til l_{1}$ is dominantly $\mu$-flavored.  However, the $\tau$ Yukawa coupling makes tau production competitive with muon production, because $\neut12$ has a small, non-negligible $H_{d}^{0}$ component.
In contrast,  $\neut34$ has a {\em dominant} $H_{d}^{0}$ component, so  $\neut34$ decays are driven by the $\tau$ Yukawa coupling, leading to far more taus than muons.

As for the right-handed sleptons, the decay $\til \tau_{2} \rightarrow \til l_{1} \tau \tau$ via off-shell neutralinos is dominant over decays to $ \til l_{1} e \tau$ or $\til l_{1} \mu \tau.$  The large $\tau$ Yukawa coupling means that $\til \tau_{2}$ couples more strongly to  $\neut34 \sim \til H_{d}^{0}$ than to $\neut12 \sim \til H_{u}^{0}$.  So, the three-body decay is dominantly mediated by   $\neut34$, which preferentially produces $\tau$'s.
For the parameter point chosen, though, $\til \tau_{2}$ decays are not important for the analysis we describe below.

\subsection{Analysis}

To study the flavor-violating signals of this parameter point at the LHC, we calculate branching ratios for 2- and 3- body decays in CalcHEP (with modified vertices), and then pass these tables to  Pythia 6.4 $\cite{Sjostrand:2006za}$, which we use to simulate full SUSY production and parton showering.  We also use Pythia to generate a sample of $t {\overline t}$ background events.

We generate 96,020 SUSY events, corresponding to $100 \; \rm fb^{-1}$ of integrated luminosity. Because the gluinos are Dirac fermions, we turn off $LL$, $RR$, $LR^*$ squark-squark production. The SUSY production modes  $\til g$-$\til q_{L,R}$, $\til q_{L}$-$\til q_{R}$, $\til g$-$\til g$, and $\til q$-${\til q}^*$  account for $63.3\%$, $9.9\%$, $9.3\%$ and $7.1\%$ of the total SUSY events generated. The remaining $10.3\%$ comes from neutralino/chargino production. 

 To suppress the background from standard model processes, we keep only those events with
\begin{itemize}
\item at least two leptons with $p_{T} >$ 10 GeV and $|\eta| <$ 2.4
\item $\sum p_{T} >$ 1500 GeV,
\end{itemize}
where $\sum p_{T}$ includes jets with $p_{T} >$ 20 GeV, leptons with $p_{T} >$ 10 GeV and $|\eta| <$ 2.4, and photons with    $p_{T} >$ 10 GeV.  After these cuts, the leading SM background is $t \bar{t}.$ The number of $t\bar{t}$ events that pass these cuts after $100 \; \rm fb^{-1}$ of integrated luminosity is $997$, versus $20,173$ events from SUSY production. 

The analysis for $e$-$\mu$ flavour violation is not qualitatively affected if the dominant production mode, $\til g$-$\til q_{L,R}$, becomes unavailable, say in the case of a very heavy gluino. For a $2$ TeV gluino, the number of SUSY events generated is $20,650$ and the number of SUSY events that pass all cuts is $2,817$. Although the signal is scaled down, the flavor-subtraction techniques employed efficiently reduce the $t\bar{t}$ background and a flavor-violating signal is still observable.

The distributions of  dilepton invariant masses have various associated endpoints. The kinematic endpoint for the dileptons in the type of decay chain  shown in Figure \ref{LFVdecay} ($\til \chi^{0} \rightarrow \til {l_1} l \rightarrow \til \nu \bar{\nu} l$) is
\begin{equation}
m_{ll}^{max} =\sqrt{\frac{m_w(m_{\til \chi^{0}}^{2}-m_{\til l}^{2})}{m_{\til l}} \sqrt{\frac{m_{\til l}^2-m_{\til \nu}^{2}+m_w^2+\sqrt{(m_{\til l}^2-m_{\til \nu}^{2}+m_w^2)^2-(2 m_{\til l} m_w )^2}}{m_{\til l}^2-m_{\til \nu}^{2}+m_w^2-\sqrt{(m_{\til l}^2-m_{\til \nu}^{2}+m_w^2)^2-(2 m_{\til l} m_w )^2}}}}.
\end{equation}
The kinematic endpoint for the taus in the decay chain $\til \chi^{0} \rightarrow \til {\tau_2} l \rightarrow \til \nu \bar{\nu} \tau$ has a simpler expression, as it involves one two-body decay and one-three body decay, rather than three two-body decays:
\begin{equation}
m_{\tau \tau}^{max} =\sqrt{\frac{(m_{\til \chi^{0}}^{2}-m_{\til \tau}^{2})(m_{\til \tau}^{2}-m_{\til \nu}^{2})}{m_{\til \tau}^{2}}}
\end{equation}

Numerical values of endpoints associated with cascades involving  $\til l_{1}$ and $\til \tau_{2}$ are given  in Table \ref{table:endpoints}.

\begin{table}
\centering
\begin{tabular}{|c|c|}
\hline
Decay Channel & $m_{ll}^{max}$ or $m_{\tau \tau}^{max}$ (GeV) \\
\hline
$\neut12 \rightarrow \til l_{1} \rightarrow \snu1 $ & 257 \\
\hline
$\neut34 \rightarrow \til l_{1} \rightarrow \snu1$ & 269 \\
\hline
\hline
$\neut12 \rightarrow \til \tau_{2} \rightarrow \snu2 $ & 207 \\
\hline
$\neut34 \rightarrow \til \tau_{2} \rightarrow \snu2 $ & 222 \\
\hline
\end{tabular}
\caption{Kinematic endpoints for decays via $\til l_{1}$ and $\til \tau_{2}$.}
\label{table:endpoints}
\end{table}

We will focus on $ee$ $\mu \mu$ and $e \mu$ opposite-sign dilepton pairs, which singles out the $\neut12 \rightarrow \til l_{1} \rightarrow \snu1$ decay chain ($\neut34 \rightarrow \til l_{1} \rightarrow \snu1$ occurs much less often). Taus produced in $\til \tau_{2}$ chains could potentially contaminate the $e$-$\mu$ signal that interests us when  they  decay leptonically, but these decays are flavor universal and are effectively subtracted out when we consider distributions that are sensitive to $e$-$\mu$ flavor violation. In our analysis, we include the effect of the decays of $\tau$s that are produced in the SUSY decay chains and $t\bar{t}$ background.

A number of strategies can be applied to gather evidence for  $e$-$\mu$ violation.  The probabilities of having an electron or a muon come from the first vertex in the decay chain of Figure \ref{LFVdecay} are directly proportional, respectively, to the $e$ and $\mu$ contents of $\til l_{1}.$ Because electrons and muons are produced with equal probability by the $W$-boson in the chain, only if the $e$-$\mu$ mixing in $\til l_{1}$ is $1:1$ will the number of $ee$ and $\mu\mu$ events be equal on average.  By considering the ratios of the numbers of $ee$, $e\mu$, and $\mu\mu$ events, information about the flavor violating slepton vertex can be extracted.

Event counts for opposite-sign $ee$, $e\mu$, and $\mu\mu$ pairs with invariant masses in the range 180 GeV $\lesssim$ $m_{ll}$ $\lesssim$ 290 GeV are given in Table \ref{table:eventcount}. These counts suggest that the lightest left-handed slepton has a larger $\mu$ than $e$ component. The mass range is chosen to ensure that the events arise predominantly from $\til l_{1}$ decays.

\begin{table}[h]
\centering
\begin{tabular}{|c|c|}
\hline
Distribution & Multiplicity \\
\hline
$ee$ & 216 \\
\hline
$\mu\mu$ & 1,711 \\
\hline
$ e\mu $ & 1,601 \\
\hline
\end{tabular}
\caption{The number of opposite-sign $ee$, $\mu\mu$ and $e\mu$ events in the mass range 180 GeV $\lesssim$ $m_{ll}$ $\lesssim$ 290 GeV.}
\label{table:eventcount}
\end{table}

\begin{figure}
\begin{center}
\includegraphics[width = 0.60\textwidth]{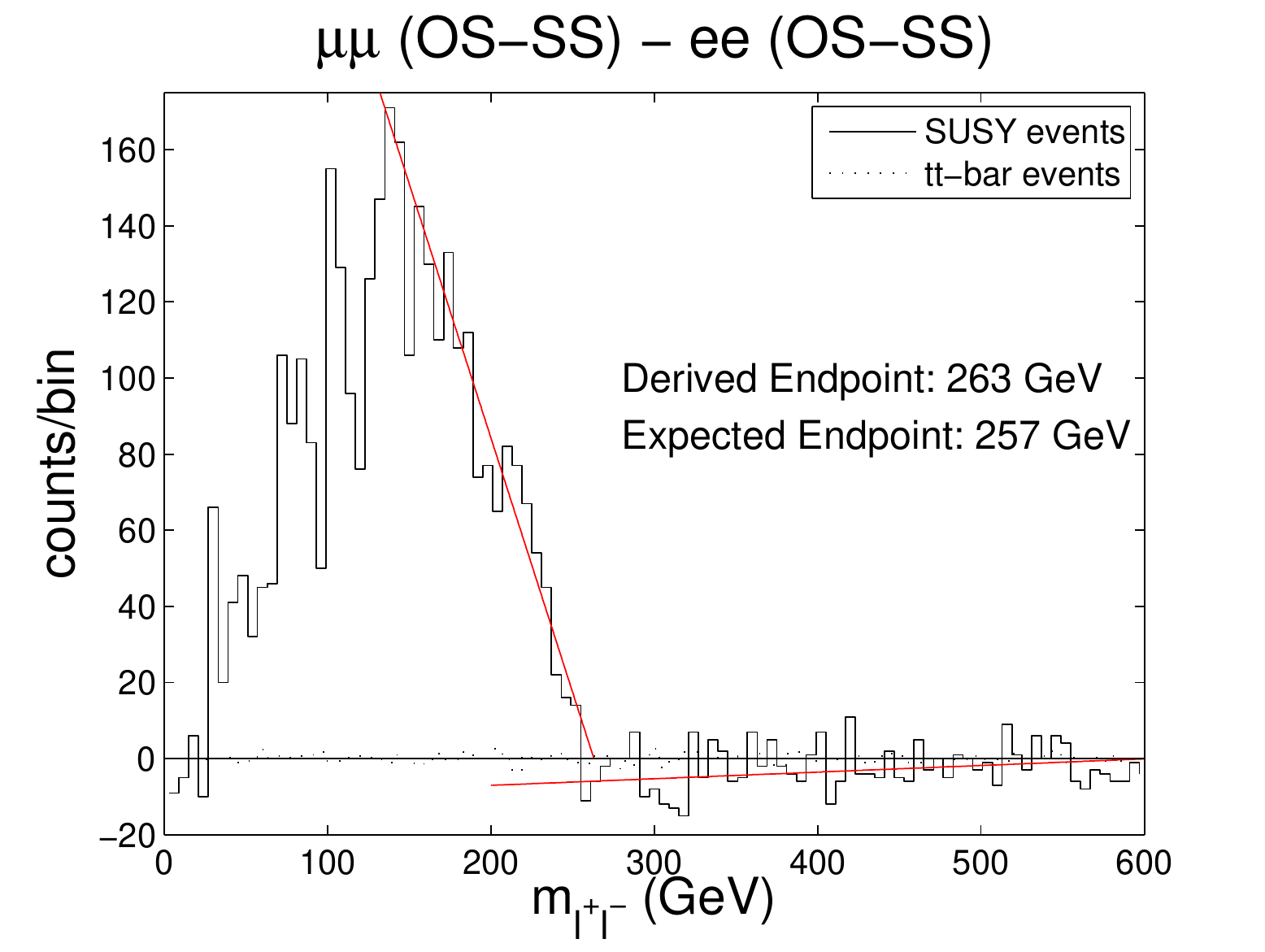}
\caption{An excess of $\mu\mu$ over $ee$ events indicates $e$-$\mu$ flavor violation}
\label{mmee}
\end{center}
\end{figure}

Subtracting $ee (OS-SS)$ invariant mass distribution from the $\mu\mu (OS-SS)$ invariant mass distribution (Fig. \ref{mmee}) not only shows that there are more $\mu$ than $e$ events, but also reveals a clear endpoint near the expected value. The $t\bar{t}$ background is suppressed by the flavor subtraction  and the $SS$ subtraction reduces combinatorial background in the SUSY events. The derived endpoint is calculated by a 2-line fit on the data, as depicted by the straight lines in the plot.

\begin{figure}
\begin{center}
\includegraphics[width = 0.60\textwidth]{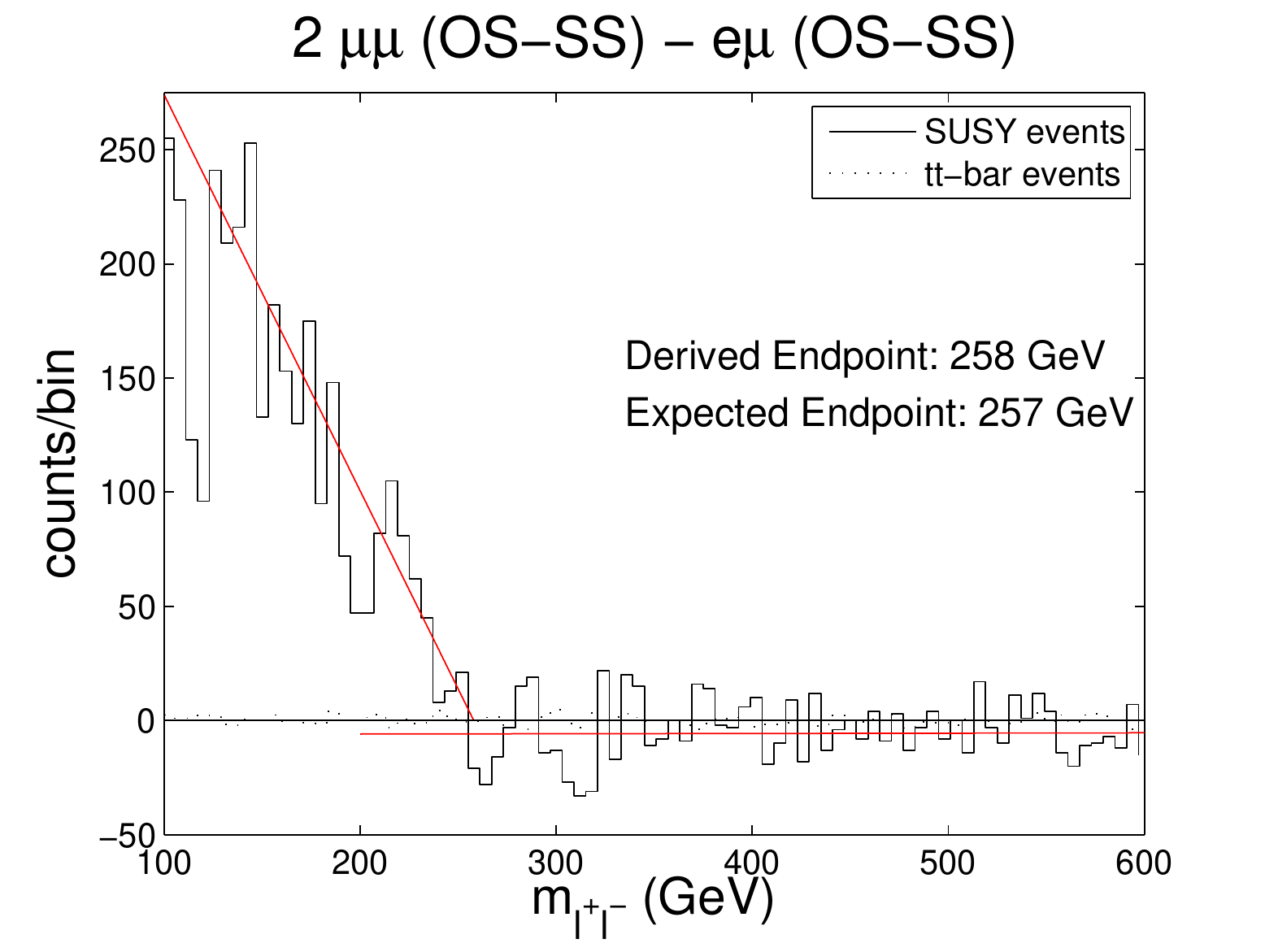}
\caption{The ratio of $\mu\mu$ to $e\mu$ events is not 1:2, suggesting $e$-$\mu$ flavor violation}
\label{mmem}
\end{center}
\end{figure}

In the absence of flavor violation, there should be twice as many $e\mu (OS)$ events as $\mu^{+}\mu^{-}$ events.  An excess after subtracting $e\mu$ events from twice the number of $\mu\mu$ events is therefore another indicator $e$-$\mu$ flavor violation. In
Table \ref{table:eventcount} we see that the numbers of $\mu^{+}\mu^{-}$ and $e\mu (OS)$ events are almost the same, rather than different by a factor of two, and  the plot in Fig. \ref{mmem} shows the relevant invariant mass distribution. The $SS$ subtraction again reduces combinatorial background in the $\mu^{+}\mu^{-}$ and $e\mu (OS)$ SUSY events.

The $e$-$\mu$ flavor violation can also be seen in the $e\mu$ events alone.
In the  decay chain of Figure \ref{LFVdecay}, the lepton from the neutralino-slepton vertex typically has a higher $p_{T}$ than the lepton arising from $W$ decay. The flavor distribution of the harder lepton can thus be used to learn about the flavor content of  $\til l_{1}$.

From the total sample of 20,173 events that pass all $p_{T}$ cuts and contain at least two leptons, the numbers of events whose highest-$p_{T}$ lepton is of $e$ or $\mu$ type are given in Table \ref{table:highestpT}. Since this is an inclusive study of all possible SUSY decay chains,  exact mixing angles cannot be derived from this analysis. However, the numbers clearly indicate that hard muons are favored over hard electrons.

\begin{table}[h]
\centering
\begin{tabular}{|c|c|}
\hline
Highest $p_{T}$ lepton & Multiplicity \\
\hline
$e$ & 5,391 \\
\hline
$\mu$ & 14,782 \\
\hline
\end{tabular}
\caption{The numbers of events passing cuts with the hardest lepton being $e$ or $\mu$.}
\label{table:highestpT}
\end{table}

\begin{figure}
\begin{center}
\includegraphics[width = 0.6\textwidth]{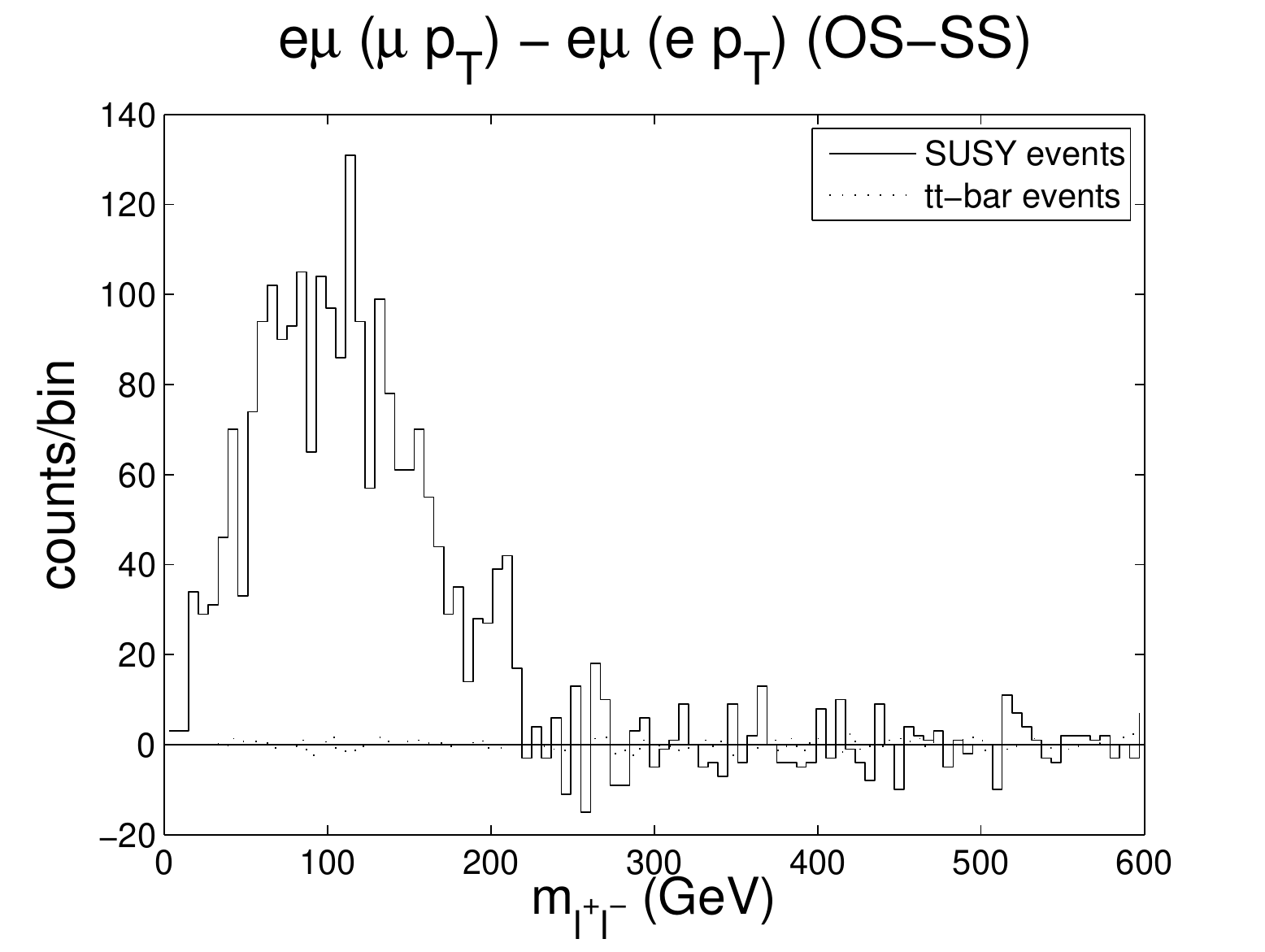}
\caption{$e\mu$ distributions for events with highest $p_{T}$ $e$ subtracted from highest $p_{T}$ $\mu$ events.}
\label{empT}
\end{center}
\end{figure}

By subtracting the dilepton invariant mass distribution for $e\mu$ events in which $e$ is the hardest lepton from the distribution for $e\mu$ events in which $\mu$ is the hardest lepton (Fig. \ref{empT}), a clear event excess indicating $e$-$\mu$ flavor violation can be observed.

That the derived endpoint does not match the endpoints seen in the previous plots is not surprising given that the bias toward harder muons does not persist throughout all regions of phase space.  For example,  if the sequence of two-body decays depicted in Figure \ref{LFVdecay} proceeds in a way that maximizes the invariant mass of the two charged leptons, then for the parameters we've adopted it turns out that the lepton from the first vertex (the one more likely to be a muon than an electron) is in fact slightly {\em less} energetic that the lepton from the second vertex, in the rest frame of the neutralino.

%Although we have focused on events with $ee$, $\mu\mu$, or $e\mu$ pairs, the $e$-$\mu$ flavor violation can in principle also be observed in events with taus.  A kinematic endpoint consistent with the ones already obtained can be extracted from the flavor-subtracted $\mu\tau (OS-SS)$-$e\tau (OS-SS)$ distribution (Fig. \ref{mtet}).

%
%
%\begin{figure}[h]
%\begin{center}
%\includegraphics[width = 0.60\textwidth]{mtOSSSetOSSS.pdf}
%\caption{Kinematic endpoint in the $\mu\tau$ - $e\tau$ OS-SS distribution}
%\label{mtet}
%\end{center}
%\end{figure}

A more sophisticated analysis might  improve discrimination of signal and background and  the readability of kinematic endpoints.  It would also be interesting to employ some level of detector simulation to explore how the types of analyses we have described might work in a real detector environment.  In particular, the different efficiencies for detecting electrons and muons may complicate the flavor subtraction technique we have used to isolate the flavor-violating signal.

\section{Conclusions}
Dark matter remains one of the most exciting topics in particle physics, and one in which some of the most significant experimental progress will be made over the next few years. We have revisited the topic of mixed sneutrino dark matter in the context of the Minimal R-symmetric Supersymmetric Standard Model (MRSSM),
focusing in particular on the connections between neutrino masses and mixings, dark matter, and collider phenomenology in this framework.

Because flavor violation is expected to be large in the MRSSM, the anarchic generation of large neutrino mixing angles finds a natural home in this theory. The small Majorana gaugino masses allow the radiatively generated neutrino masses to be of the appropriate size, while still allowing enough lepton-number violation for the LSP sneutrino to evade strong direct detection constraints on scattering via $Z$-exchange.
We find that the region of neutrino parameter space defined by the current experimental constraints is consistent with the idea that  the neutrino mass matrix is generated by an essentially random set of sneutrino mass matrices.

Data from neutrino experiments give some insight into the expected properties of the LSP sneutrino. First, the LSP tends to be dominantly sterile. This is to the benefit of the model, as this same region of parameter space is preferred by the measured relic abundance. Second, the LSP tends to have significant $\mu$ and $\tau$ components. The consequence of a large $\mu$ component in the LSP is that studies of $\mu \mu$, $ee$, and $e \mu$ invariant-mass distributions can give conclusive evidence of
lepton flavor violation
at the LHC. In particular, inclusive signatures of opposite sign dileptons can reveal strong flavor violation that would not appear in the MSSM, even with small non-degeneracies in the first two generations.

Direct detection experiments have a great possibility of probing this model. While the model provides a natural basis for inelastic dark matter, it also predicts a sizeable elastic scattering cross section ($\gsim 10^{-44} {\rm cm^2}$) and should give detectable signals in the next generation of dark matter direct detection experiments. Should such a scenario be discovered, we may be able to probe the origin of the mysterious properties of the neutrino mass matrix at the weak scale.

\acknowledgments{The authors thank Patrick Fox for many useful discussions. We thank Graham Kribs and Ricky Fox for pointing out an error in the 
previous version of this paper. NW and AK are supported by NSF CAREER grant PHY-0449818 and DOE OJI grant \# DE-FG02-06ER41417. DTS is supported by NSF grant PHY-0856522.}

\bibliographystyle{JHEP}
\bibliography{Sneutrinos}

\end{document}